\documentclass{ws-book961x669}
\usepackage{ws-book-thm}     
\usepackage{ws-book-har}    
\usepackage[pdfpagelabels=false]{hyperref}  

\begin{document}


\chapter{High--energy scattering amplitudes in QED, QCD and supergravity\label{ch1}}

\begin{center}
{\bf \large Agust{\' \i}n Sabio Vera}\\ 
\vspace{.4cm}
{\small CERN, Theoretical Physics Department, Geneva, Switzerland.}\\
{\small Instituto de F{\' \i}sica Te{\' o}rica UAM/CSIC, Nicol{\'a}s Cabrera 15}\\ 
{\small \& Universidad Aut{\' o}noma de Madrid, E-28049 Madrid, Spain.}
\end{center}

\section{Lipatov's high energy effective action for gravity}

Phenomenology at the LHC and the AdS/CFT correspondence~\cite{Maldacena:1997re,Gubser:1998bc,Witten:1998qj} have motivated a plethora of new results related to the calculation of scattering amplitudes. This is, in particular, the case for  
${\cal N} = 4$ supersymmetric Yang--Mills theory~\cite{Bern:2010tq} and ${\cal N} = 8$ supergravity~\cite{Bern:1998ug,Bern:2009kd}. An important finding in this context was the double--copy structure connecting both theories~\cite{Bern:2002kj}.

An interesting limit in graviton scattering at high center--of--mass energy, $s$,  is that of multi--Regge kinematics,  where the amplitudes factorize, with reggeized gravitons exchanged in the $t$--channel~\cite{Grisaru:1975tb,Grisaru:1981ra}. There are also contributions related to eikonal and double--logarithmic in $s$ (DL) terms~\cite{Lipatov:1982vv,Lipatov:1982it,Lipatov:1991nf} which can be calculated using the high energy effective action proposed by Lipatov~\cite{Lipatov:2011ab}. Following Gribov's proposal for the construction of an effective quantum field theory for Pomeron interactions, Lipatov worked on building it for gravity. The basic idea underlying this approach is that gravitons are not elementary particles when radiative corrections are taken into account, they lie on their Regge trajectories. 

Lipatov's effective action for gravity shares common ground with the Quantum Chromodynamics  (QCD) case. It is based on the concept that the leading in energy contributions to the cross section, besides eikonal contributions, are those with gravitons produced at bunches well separated in rapidity from each other. Each gap in rapidity between subsequent bunches is generated by the exchange of a reggeized graviton. There exist generally covariant fields which create, $A^{++}$, or annihilate, $A^{--}$, them in the $t$--channel and are subject to kinematical constraints $\partial_+ A^{++} = \partial_- A^{--} =0 $ to create the rapidity gaps. Within a bunch there appear local and non--local interactions described by the action~\cite{Lipatov:2011ab}
\begin{eqnarray}
S = -\frac{1}{2 \kappa} \int d^4 x \bigg(\sqrt{-g} R + \partial_\sigma A^{++} 
\partial_\sigma A^{--} + \frac{1}{2} \left(\partial_+ j^- 
\partial^2_\mu A^{++} + [+ \leftrightarrow -] \right) \bigg)
\end{eqnarray}
where $\kappa^2 = 8 \pi G = 8 \pi^2 \alpha$, and $G$ is  Newton's constant. It corresponds to the Einstein--Hilbert action plus a kinetic term for the reggeon fields and induced contributions. In the latter, the currents are related to the usual graviton field $h_{\mu \nu}$ in the form
\begin{eqnarray}
\partial_\pm j^\mp &=& h_{\pm \pm} - \left(h_{\rho \pm} - \frac{1}{2}
\frac{\partial_\rho}{\partial_\pm} h_{\pm \pm}\right)^2 + \dots
\end{eqnarray}
This accounts for the interactions among usual gravitons and reggeized ones when inserted back in the action. 

From this approach it is possible to obtain the graviton Regge trajectory and effective vertices needed to calculate elastic and  inelastic amplitudes at any order in the theory. It is a compact formalism which allows for the resummation of the leading contributions from an infinite number of Feynman diagrams. In particular, the non--local $1/\partial_\pm$ operators generate graviton emissions from hard lines far in rapidity from the local clusters. In order to preserve general covariance, Lipatov's high energy effective action also contains eikonal contributions. 

In relation with the underlying double--copy structure of gravity,  it is worth noting that the graviton emission vertex in multi--Regge kinematics is directly related to the equivalent~\cite{BFKL1,BFKL2,BFKL3} gluon emission vertex in QCD~\cite{Lipatov:1982vv,Lipatov:1982it,Lipatov:1991nf,SabioVera:2011wy} as it was originally found by Lipatov.
 
\section{Double logarithms in supergravity}

There exists a subtle interplay between the reggeization of the graviton and the DL contributions to the scattering amplitudes. In~\cite{Bartels:2012ra} these DLs in Einstein--Hilbert gravity and different supergravities were calculated to all orders.  The results there presented are in agreement up to two loops with those obtained for ${\cal N}=4,5,6$ supergravities using a double--copy procedure~\cite{BoucherVeronneau:2011qv}.  They also agree with a recent three loop calculation for ${\cal N}= 8$  supergravity~\cite{Henn:2019rgj}. A remarkable feature of the all--orders resummation in the DL sector of the amplitudes is that, when compared to their Born counterparts, they grow with $s$ whenever ${\cal N}<4$ but rapidly decrease with $s$ if ${\cal N}>4$. 

Let us take the normalizations ($L$=loop order, $N$=number of gravitinos) for the 
 4--graviton scattering amplitude with helicities (++;++),  
\begin{eqnarray}
{\cal A}_{4,(N)} &=& {\cal A}_4^{\rm Born} {\cal M}_{4,(N)}
~=~ \kappa^2 \frac{s^3}{t u} \left(1 + \sum_{L=1}^\infty {\cal M}_{4,(N)}^{(L)}\right),\label{eqnnotation}
\end{eqnarray}
where $s=(p_1+p_2)^2$, $t=(p_1-p_3)^2$ and $u=(p_1-p_4)^2$. Lipatov found that the Regge limit, $s \gg -t={\vec{q}}^{\, 2}$,  of the one loop amplitude contains the graviton Regge trajectory~\cite{Lipatov:1982vv,Lipatov:1982it,Lipatov:1991nf}, 
\begin{eqnarray}
\omega (q) &=& \frac{\alpha \, {\vec{q}}^{\, 2}}{\pi}\int \frac{d^2 \vec{k} }{\vec{k}^2 
(\vec{q}-\vec{k})^2}\nonumber\\
&\times&  \left((\vec{k} \cdot (\vec{q}-\vec{k}))^2\left(\frac{1}{\vec{k}^2 }+
\frac{1}{(\vec{q}-\vec{k})^2}\right)+\frac{N}{2}\,(\vec{k} \cdot (\vec{q}-\vec{k}))-{\vec{q}}^{\, 2}\right).
\end{eqnarray}
Its infrared divergence is regularized by a cut--off, $\lambda$, while the ultraviolet one, which is due to the use of an effective action at high energies, can be regularized by $s$, taken as an upper cut--off. One then gets
\begin{equation}
\omega (q) = \alpha \,t 
\,\left(\ln \left(\frac{-t}{\lambda ^2}\right) +\frac{N-4}{2}\,\ln \left(\frac{s}{-t}\right)\right)\,.
\label{trajgrav}
\end{equation}
A similar structure remains to all orders in perturbation theory. With DL accuracy, it is useful to use the representation
\begin{equation}
{\cal A}_{4,(N)}  = {\cal A}_4^{\rm Born}\,\left(\frac{s}{-t}\right)^{\alpha t\,\ln \left(\frac{-t}{\lambda ^2}\right)}
\int _{\delta-i\infty}^{\delta +i\infty}\frac{d\,\omega }{2\pi i}\,
\left(\frac{s}{- t}\right)^\omega \frac{ f ^{(N)}_\omega}{\omega} \,,\,\,\delta>0\, ,
\label{factor}
\end{equation}
where the $t$--channel partial wave $f ^{(N)}_\omega$ has the expansion
\begin{equation}
f^{(N)}_\omega =\sum _{n=0}^\infty {\cal C}_n^{(N)}\,\left(\frac{\alpha t}{\omega^2}\right)^n\,.
\label{fomegaexp}
\end{equation}

In~\cite{Lipatov:1982vv} Lipatov calculated the DL contributions to the 4--graviton amplitude stemming from ladder diagrams. In~\cite{Bartels:2012ra} it was shown that non--ladder contributions should also be taken into account. These extra terms already appear in quantum electrodynamics (QED) where, {\it e.g.}, in $e^+e^-$ forward scattering the scattering amplitude factorizes at 
DL accuracy: ${\cal A}_{\rm DL} (s) = {\cal A}_{\rm Born} {\cal R} (s)$. An infrared cut--off $\mu$ can be introduced for the transverse momenta of the virtual contributions, ${\vec{p}_T}^{\, 2} > \mu^2$. It is then possible to write ${\cal R} (s) \to {\cal R} (s,\mu^2)$ and the following equation holds in Sudakov variables,
\begin{eqnarray}
{\cal R} (s,\mu^2) &=& 1 \nonumber\\
&&\hspace{-1.5cm}+ \,  e^2 \int_{-\infty}^\infty \frac{d (s \alpha)}{s \alpha}
\int_{-\infty}^\infty  \frac{d (s \beta)}{s \beta} \int \frac{d^2 \vec{k}}{i (2 \pi)^4} 
\frac{{\vec{k}}^{2} \, \theta ({\vec{k}}^{2} - \mu^2)}{(s \alpha \beta - {\vec{k}}^{2})^2} {\cal R} (s \alpha,{\vec{k}}^{2})
{\cal R} (s \beta,{\vec{k}}^{2}) .
\end{eqnarray}
Its solution can be expressed in terms of a Bessel function. For $e^+e^-$ backward scattering the contributions from Sudakov photon lines attached to the external fermions kick in, and the corresponding equation has a solution 
written in terms of a parabolic cylinder function which will be described below. 

These are classical results which were obtained by Gorshkov, Gribov, Lipatov and Frolov from 1966 to 1970~\cite{Gorshkov:1966ht,Gorshkov:1966qd,Gorshkov:1966hu}. In QCD, in deep inelastic electron--hadron scattering at large photon virtualities, one can modify the usual DGLAP~\cite{Gribov:1972ri,Dokshitzer:1977sg,Altarelli:1977zs} evolution equation for parton distributions introducing a dependence mixing longitudinal and transverse degrees of freedom to generate the DL contributions~\cite{Kirschner:1982qf,Kirschner:1982xw,Kirschner:1983di}. Similar techniques have been applied to the electroweak sector of the Standard Model~\cite{Fadin:1999bq}.

Coming back to graviton scattering and introducing the gravitino content of supergravity, at one loop the result is given in Eq.~(\ref{trajgrav}). This can be compared with the exact amplitude in, {\it e.g.}, ${\cal N}=8$ supergravity (where $N=8$),
\begin{eqnarray}
{\cal M}^{(1)}_{4, (N=8)}  &=& \underbrace{\alpha \, t \ln{\left(\frac{-s}{-t}\right)}\ln{\left(\frac{-u}{-t}\right)}}_{\rm Double~Logs} - \underbrace{\alpha \frac{(s-u)}{2} \ln{\left(\frac{-t}{\lambda^2}\right)}\ln{\left(\frac{-s}{-u}\right)}}_{\rm Eikonal}\nonumber\\
&+& \underbrace{\alpha \, \frac{t}{2} \ln{\left(\frac{-t}{\lambda^2}\right)}
\left(\ln{\left(\frac{-s}{-t}\right)}+\ln{\left(\frac{-u}{-t}\right)}\right)}_{\rm Trajectory}.
\label{Mamplitude}
\end{eqnarray}
There are three contributions: DLs, the eikonal piece and the graviton Regge trajectory. Corrections to this formula at two loops for different supergravities have been given in~\cite{BoucherVeronneau:2011qv} and at one loop in Einstein--Hilbert gravity ($N=0$) in~\cite{Dunbar:1994bn}. They all agree with Eq.~(\ref{trajgrav}) at one--loop order. Its generalization  to all orders can be written as the solution of an evolution equation for the partial wave in Eq.~(\ref{fomegaexp}) which is similar to those obtained in QED and QCD, {\it i.e.}
\begin{equation}
f^{(N)}_\omega =1 - \alpha \, t \frac{d}{d\,\omega}\,\frac{f^{(N)}_\omega}{\omega}
+ \alpha \, t \frac{(N-6)}{2} \frac{{f^{(N)}_\omega}^2}{\omega ^2}.
\label{evoleq}
\end{equation}
The derivative term is due to virtual gravitons with the smallest $p_T$. Gravitons and gravitinos exchanged in the $t$--channel with very low energy generate the quadratic contribution. The perturbative solution can be obtained by iteration,
\begin{eqnarray}
f_\omega^{(N)} &=& 1 + \frac{\alpha \, t (N-4)}{2 w^2} + \frac{\alpha^2 t^2 (N-4) (N-3)}{2 w^4} \nonumber\\
    &+& \frac{\alpha^3 t^3 (N-4) \left(5 N^2- 26 N+36\right)}{8 w^6}    + \dots
   \label{solveq}
   \end{eqnarray}
Going back to the DL amplitude, this implies,
\begin{equation}
{\cal A}_{4,(N)} = {\cal A}_4^{\rm Born} \left(\frac{s}{-t}\right)^{\alpha \, t\,\ln 
\left(\frac{-t}{\lambda ^2}\right)}\,\Phi^{(N)} \left(- \alpha \, t \,\ln^2 \left(\frac{s}{-t}\right)\right)\,,
\end{equation}
where
\begin{eqnarray}
\Phi^{(N)} (x ) &=&1-\frac{(N-4)}{2}\,\frac{x}{2}+\frac{(N-4)}{2}(N-3)\frac{x^2 }{4!}\nonumber\\
    &-&\frac{(N-4)}{8}(5N^2-26N+36)\frac{x ^3}{6!}    + \dots
\label{ampexpand}
\end{eqnarray}
This formula is, up to two loops for $N=4,5,6,8$, in agreement with~\cite{BoucherVeronneau:2011qv,Dunbar:1994bn} and, at three loops, for $N=8$, with~\cite{Henn:2019rgj}.

It is possible to find the exact solutions to Eq.~(\ref{evoleq}) which are compatible with the perturbative expansion in Eq.~(\ref{solveq}). These solutions were studied in~\cite{Bartels:2012ra}. The simplest one corresponds to $N=4$ since, as it can be seen in Eq.~(\ref{solveq}), only the first term survives. This implies
\begin{equation}
{\cal A}_{4,(N=4)} = {\cal A}_4^{\rm Born} \left(\frac{s}{-t}\right)^{\alpha \, t\,\ln \left(\frac{-t}{\lambda ^2}\right)}\,.
\end{equation}
The next--to--simplest case is $N=6$ because the quadratic term in the differential equation~(\ref{evoleq}) vanishes and one gets
\begin{equation}
{\cal A}_{4,(N=6)} = {\cal A}_4^{\rm Born}\,\left(\frac{s}{-t}\right)^{\alpha \, t\,\ln \left(\frac{-t}{\lambda ^2}\right)}
\,e^{\frac{\alpha \, t}{2}\ln ^2 \left(\frac{s}{-t}\right)} \, .
\end{equation}
The $N=0,2$ cases also have a simple representation:
\begin{eqnarray}
{\cal A}_{4,(N=2)} &=& \frac{{\cal A}_4^{\rm Born}}{2}
\left(\frac{s}{-t}\right)^{\alpha \, t\,\ln \left(\frac{-t}{\lambda ^2}\right)}  \left(\left(\frac{s}{-t}\right)^{\sqrt{-\alpha t}}
+\left(\frac{s}{-t}\right)^{-\sqrt{-\alpha t}}\right),\\
{\cal A}_{4,(N=0)} &=& \frac{{\cal A}_4^{\rm Born}}{3}
\left(\frac{s}{-t}\right)^{\alpha \, t\,\ln \left(\frac{-t}{\lambda ^2}\right)}  
\left(1+ \left(\frac{s}{-t}\right)^{\sqrt{-3 \alpha t}}
+\left(\frac{s}{-t}\right)^{-\sqrt{-3 \alpha t}}\right) \, .
\end{eqnarray}

The solutions with odd $N$  and $N=8$ enjoy a richer structure since their partial waves have an infinite number of poles in the complex $\omega$ plane placed asymptotically close to the lines with argument $\pm \frac{3}{4} \pi$. For these cases one can write the DL contributions to the amplitude in the form
\begin{eqnarray}
r^{(N)} (s,t) &\equiv& \frac{{\cal A}_{4,(N)} }{ {\cal A}_4^{\rm Born}} 
\,\left(\frac{s}{-t}\right)^{-\alpha \, t\,\ln \left(\frac{-t}{\lambda ^2}\right)}\nonumber\\
&&\hspace{-1.2cm}= \frac{\Gamma \left(\frac{N}{2}-3\right)}{3-\frac{N}{2}} 
\int_{\delta-i \infty}^{\delta + i \infty} \frac{d \omega}{ 2 \pi i} 
\left(\frac{s}{-t}\right)^\omega 
\frac{d}{d\,\omega }\,
\ln \left( e^{- \frac{\omega^2}{4 \alpha \, t}}D_{\frac{6-N}{2}} \left(\frac{\omega}{\sqrt{- \alpha \, t}}\right) \right)
\end{eqnarray}
with the parabolic cylinder function being $D_\nu (x) = \frac{e^{-\frac{x^2}{4}}}{\Gamma (-\nu)}\, \int_0^\infty \frac{dy}{y^{\nu +1}} \,e^{-\frac{y^{2}}{2}}e^{-x\,y}\,$.

\begin{figure}
\begin{center}
\includegraphics[width=11.cm]{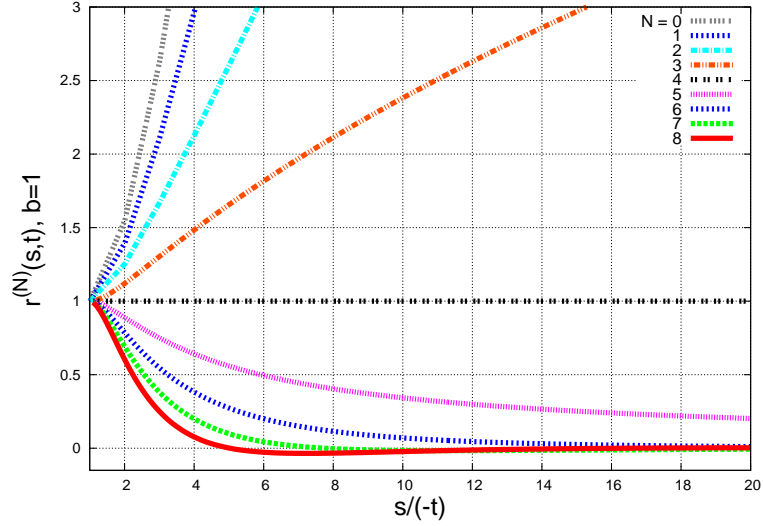}
\vspace{-1.3cm}
\end{center}
\caption{Scattering amplitude for $N=0,1, \dots,8$.}
\label{rAllb1}
\end{figure}

It is now possible to numerically compare the behaviour with $s$ of the amplitudes for different $N$. Taking $b \equiv - \alpha \, t = 1$ one can see the result in Fig.~\ref{rAllb1}. 
The solutions for $N<4$ grow monotonically with energy. The critical solution with $N=4$ is constant and equal to one. There are two monotonically decreasing with energy solutions for $N=5,6$ and, finally, two oscillatory and decreasing with energy ones for $N=7,8$.  The high energy behaviour is driven by the poles with the largest real part, whose values can be evaluated numerically. This allows to obtain asymptotic representations for different $N$:
\begin{eqnarray}
r^{(N=0)}(s,t) &\approx& s^{\sqrt{-3 \alpha \, t}}\,,\\
r^{(N=1)}(s,t) &\approx&s^{1.402\,\sqrt{- \alpha \, t}}\,, \\
r^{(N=2)}(s,t) &\approx& s^{\sqrt{- \alpha \, t}}\,,\\
r^{(N=3)}(s,t) &\approx& s^{0.5508\,\sqrt{- \alpha \, t}}\,,\\
r^{(N=4)}(s,t) &=& 1\,,\\
r^{(N=5)}(s,t) &\approx& s^{-0.762\,\sqrt{- \alpha \, t}}\,,\\
r^{(N=6)}(s,t) &=& e^{\frac{\alpha}{2} t \ln^2 s}\,,
\end{eqnarray}
and
\begin{eqnarray}
r^{(N=7)}(s,t) &\approx&
s^{-2.111 \,\sqrt{- \alpha \, t}}\cos \left(2.267 \,\sqrt{- \alpha \, t}
\ln s\right)\,,\\
r^{(N=8)}(s,t) &\approx&
s^{-1.916 \,\sqrt{- \alpha \, t}}\cos \left(2.816 \,\sqrt{- \alpha \, t}
\ln s\right)\,.
\label{r8asymp}
\end{eqnarray}

\section{Impact parameter representation}

In order to gain further understanding for the resummation of DLs, it is useful to work in impact parameter ($\rho$) representation. This has been done for ${\cal N} \geq 4$ supergravities in~\cite{SabioVera:2019edr,SabioVera:2019jqe} where  a novel representation for the coefficients of the perturbative expansion is introduced using mathematical results on exactly solvable recurrences.  In this representation it can be shown  that the DLs make the gravitational interaction weaker at small $\rho$ and large $s$. This is so because they   introduce corrections to the Born eikonal phase which  change the sign of the graviton's deflection angle.  As a consequence, there is a constant negative shift in both the eikonal phase and Shapiro's time delay but not large enough to generate causality violation.

The scattering amplitude can be written in the form
\begin{eqnarray}
{\cal M}_{4,{\rm DL}}^{(N)}   (s,t) = \int _{\delta-i\infty}^{\delta +i\infty}\frac{d\,\omega }{2\pi i}\,\left(\frac{s}{-t}\right)^\omega \frac{ f _\omega^{(N)} }{\omega} 
=  1+ \sum_{n=1}^\infty {\cal C}_n^{(N)} \, 
\frac{(\alpha \, t)^n}{(2n)!} \ln^{2n}{\left(\frac{s}{-t}\right)}
\label{PertAmplDL4}
\end{eqnarray} 
where the coefficients (see Eq.~(\ref{fomegaexp})) follow the relation (with $\eta_N = (N-6)/2$),
\begin{eqnarray}
 {\cal C}_n^{(N)} &=&  (\eta_N+2n-1) \, {\cal C}_{n-1}^{(N)}      +  \eta_N\sum _{j=1}^{n-1} {\cal C}_j^{(N)} {\cal C}_{n-1-j}^{(N)}  , \, \, \, {\cal C}_0^{(N)} = 1 \, . 
 \label{recursion}
\end{eqnarray}

${\cal C}^{N=4}_n = \delta_n^0$ indicating that the DLs are not present in ${\cal N}=4$ supergravity which acts as a critical theory marking the transition from poorly behaved amplitudes for $N<4$, blowing up at large $s$, to finite amplitudes when $N>4$. 

The coefficients ${\cal C}_n^{(8)}$ have already appeared in many body theory~\cite{Ihrig:1976ih}, in QED~\cite{Cvitanovic:1978wc} and in solid state physics~\cite{Goodvin:2006Go}. Eq.~(\ref{recursion}) has been studied in~\cite{MartinKearney:2011} in the context of self--convolutive recurrence relations allowing to write 
\begin{eqnarray}
{\cal C}_n^{(N)} &=& \int_0^\infty x^{n} \mu^{(N)} (x) \, dx
\label{coeffsCnN}
\end{eqnarray}
with 
\begin{eqnarray}
\mu^{(N)} (x) &=& \frac{ \frac{2^{1-\eta_N}}{\eta_N } 
\Gamma   \left(\eta _N\right)
     \sqrt{\frac{2 e^{x }}{\pi x }} }{\Gamma \left(\frac{\eta
   _N}{2}\right){}^2 \, _1F_1\left(\frac{1-\eta
   _N}{2};\frac{1}{2};\frac{x}{2}\right){}^2+2 \, x \, \Gamma \left(\frac{1+\eta _N}{2}
   \right){}^2 \, _1F_1\left(\frac{2-\eta
   _N}{2};\frac{3}{2};\frac{x}{2}\right){}^2}  
   \label{muNncoeffs} 
\end{eqnarray}
where $\Gamma$ is the gamma function and $_1F_1$ is the Kummer confluent hypergeometric function. This leads to the novel representation of the scattering amplitude
 \begin{eqnarray}
{\cal M}_{4,{\rm DL}}^{(N)}  (s,t) &=&   
 \int_0^\infty \, dx \, \mu^{(N)} (x) 
\cosh \left(  \sqrt{\alpha \, t \, x} \ln \left(\frac{s}{-t}\right)\right) \, .
\label{M4DLcosh}
\end{eqnarray} 
The $N>4$ cases have been studied in~\cite{SabioVera:2019edr,SabioVera:2019jqe} and 
read  
\begin{eqnarray}
\mu^{(5)} (x) &=& \frac{16 \sqrt{\frac{e^x}{x}}}{\Gamma\left(-\frac{1}{4}\right)^2 \, _1F_1\left(\frac{3}{4};\frac{1}{2};\frac{x}{2}\right){}^2+2 \, x \, \Gamma \left(\frac{1}{4}\right)^2 \, _1F_1\left(\frac{5}{4};\frac{3}{2};\frac{x}{2}\right){}^2} \, ,
\label{muN5}\\
\mu^{(6)} (x) &=& \frac{1}{ \sqrt{2 \pi x e^x }} \, ,
\label{muN6}\\
\mu^{(7)} (x) &=& \frac{\frac{4 \sqrt{2} }{\pi ^2 x} }{
   I_{-\frac{1}{4}}\left(\frac{x}{4}\right){}^2+I_{
   \frac{1}{4}}\left(\frac{x}{4}\right){}^2}\, ,
   \label{muN7}\\
   \mu^{(8)} (x) &=&  \frac{ \sqrt{\frac{2 e^x}{\pi^3 x}}}{\text{erfi}\left(\sqrt{\frac{x}{2}}\right
   )^2+1}\, ,
   \label{muN8}
\end{eqnarray}
where $I_{a}(z)$ is a Bessel function and ${\rm erfi} (z)$ the error function. 

To work in impact parameter space, $\rho$, the Fourier transform of the graviton--graviton interaction is introduced (with $q=\sqrt{-t}$), {\it i.e.}
\begin{eqnarray}
\chi^{(N)} \left(\rho,s\right) &=&  \alpha s 
\int \frac{d^2 \vec{q}}{q^2} e^{- i \vec{q} \cdot \vec{\rho}} 
{\cal M}_{4,{\rm DL}}^{(N)}  (s,t) \, .
\label{phaseikonal} 
\end{eqnarray}
This phase can be exponentiated in a high energy and fixed impact parameter eikonal approach,
\begin{eqnarray}
\frac{\alpha s^{2}}{q^{2}} {\cal M}_{4,{\rm DL}}^{(N)}  (s,t) 
&=& \frac{s}{4 \pi^{2}} \int d^{2} \vec{\rho} \, e^{i \vec{q} \cdot \vec{\rho}} \chi^{(N)} (\rho, s) 
\nonumber\\
&\simeq& -\frac{i s}{4 \pi^{2}} \int d^{2} \vec{\rho} \, e^{i \vec{q} \cdot \vec{\rho}}\left(e^{i \chi^{(N)} (\rho, s)}-1\right). 
\end{eqnarray}
In the forward high energy limit the integral is dominated by the region $ \frac{\partial}{\partial \rho}\left(q \rho+\chi^{(N)}\right)=0$. The corresponding graviton--graviton scattering angle reads
\begin{eqnarray}
\theta^{(N)} (\rho,s) &=& -\frac{2}{\sqrt{s}}\frac{\partial 
\chi^{(N)} \left(\rho,s\right)}{\partial \rho}   \, .
\end{eqnarray}

\begin{figure}[h]
\begin{center}
\includegraphics[width=9cm]{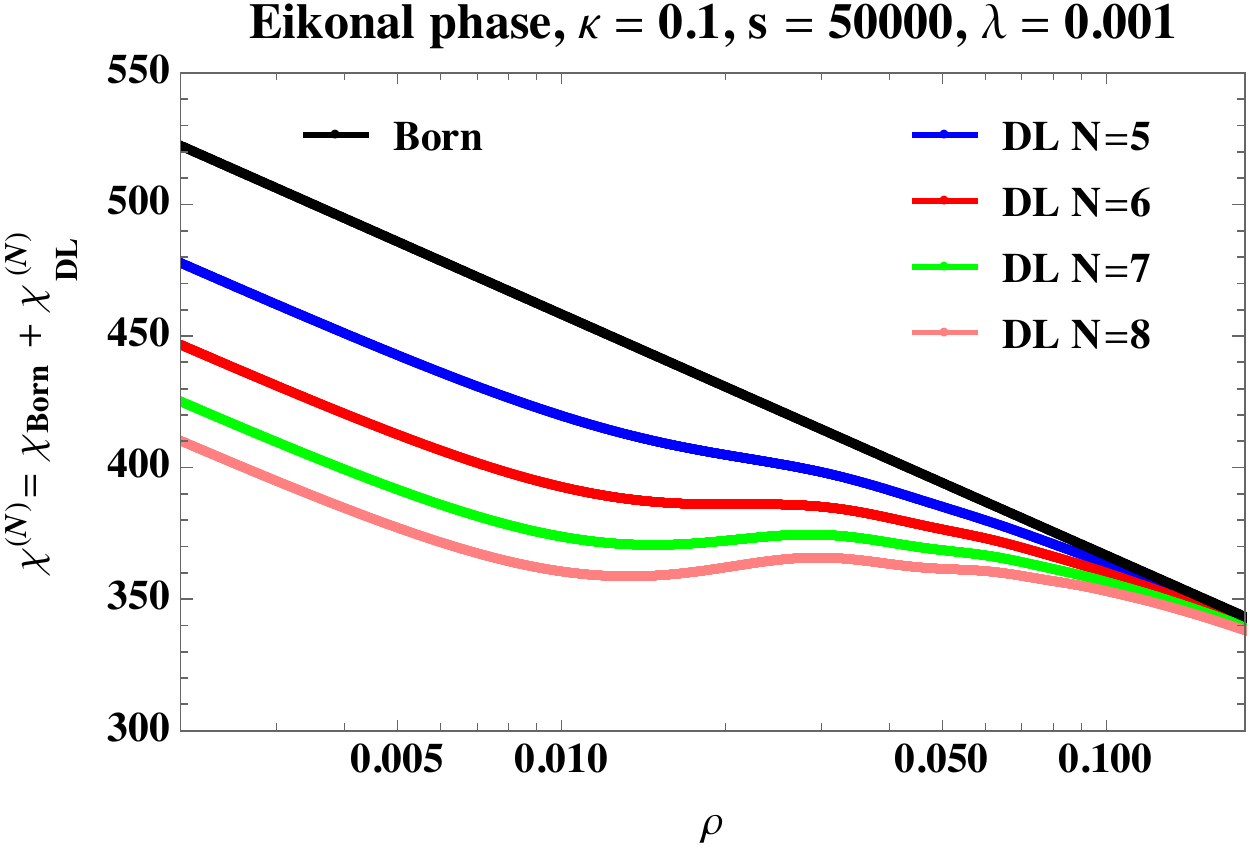}
\end{center}
\vspace{-.5cm}
\caption{Born eikonal phase plus the  DL terms for $s=50000$ and 
different supergravities. }
\label{EikonalPhaseSeveralN}
\end{figure}

\begin{figure}[h]
\begin{center}
\includegraphics[width=9cm]{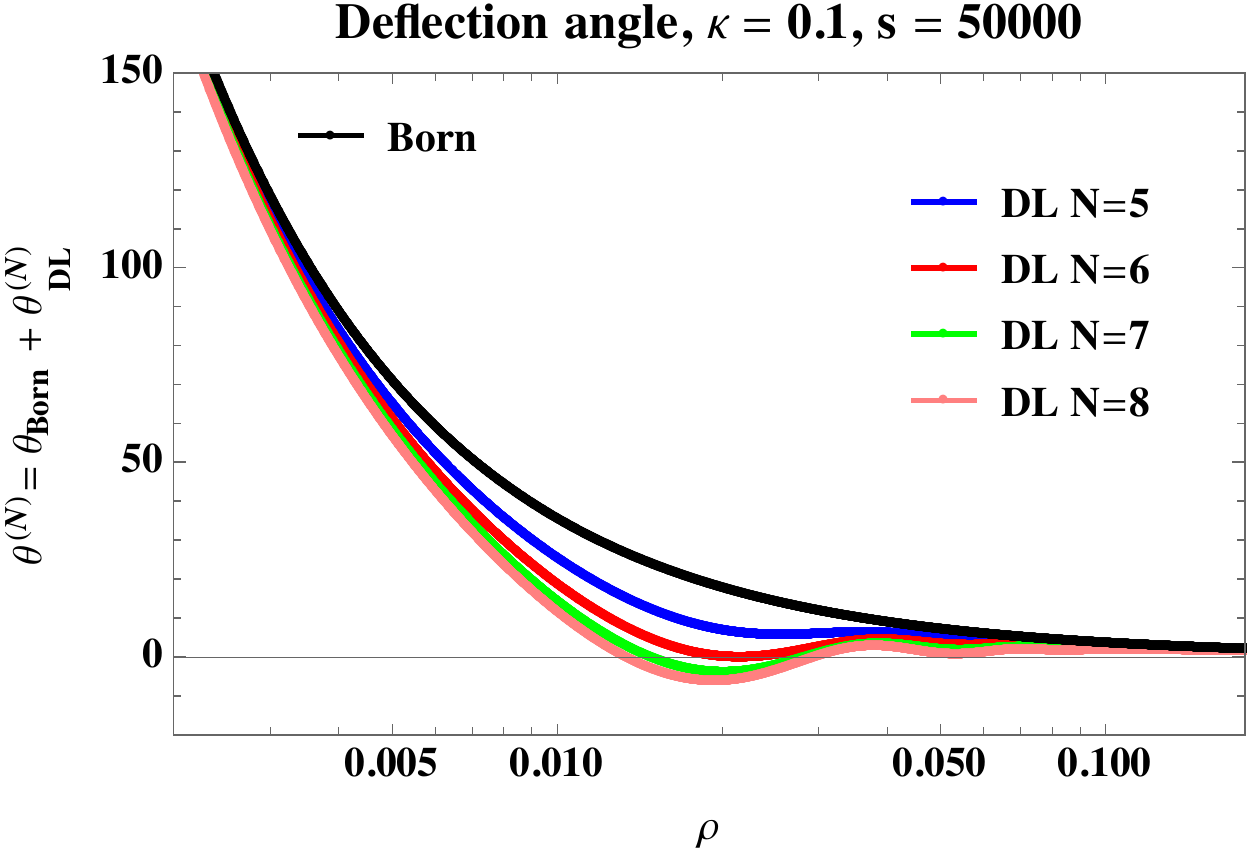}
\end{center}
\vspace{-.5cm}
\caption{Deflection angle as a function of the impact parameter for different supergravity theories.}
\label{DeflectionAngleSeveralN}
\end{figure}

\begin{figure}[h]
\begin{center}
\includegraphics[width=9cm]{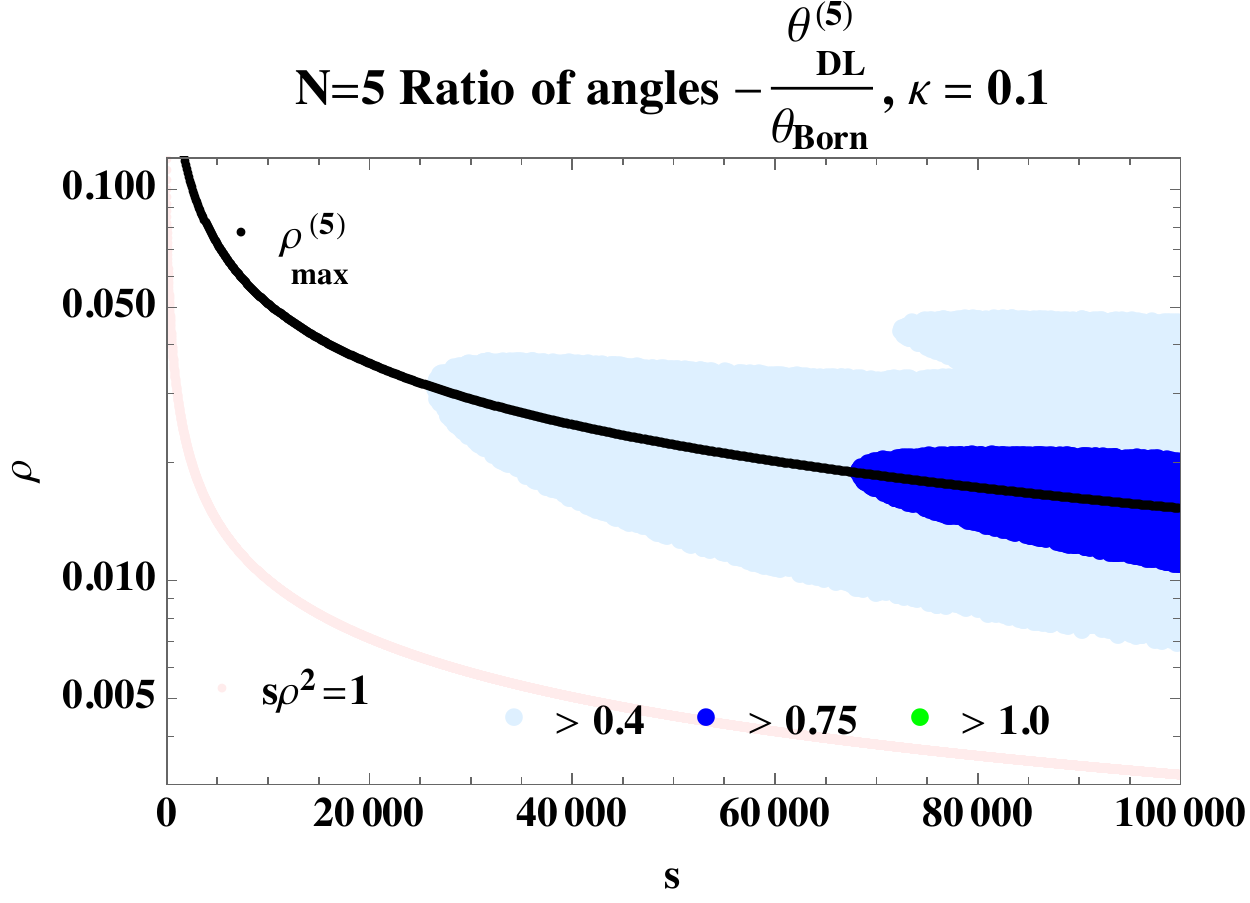}
\includegraphics[width=9cm]{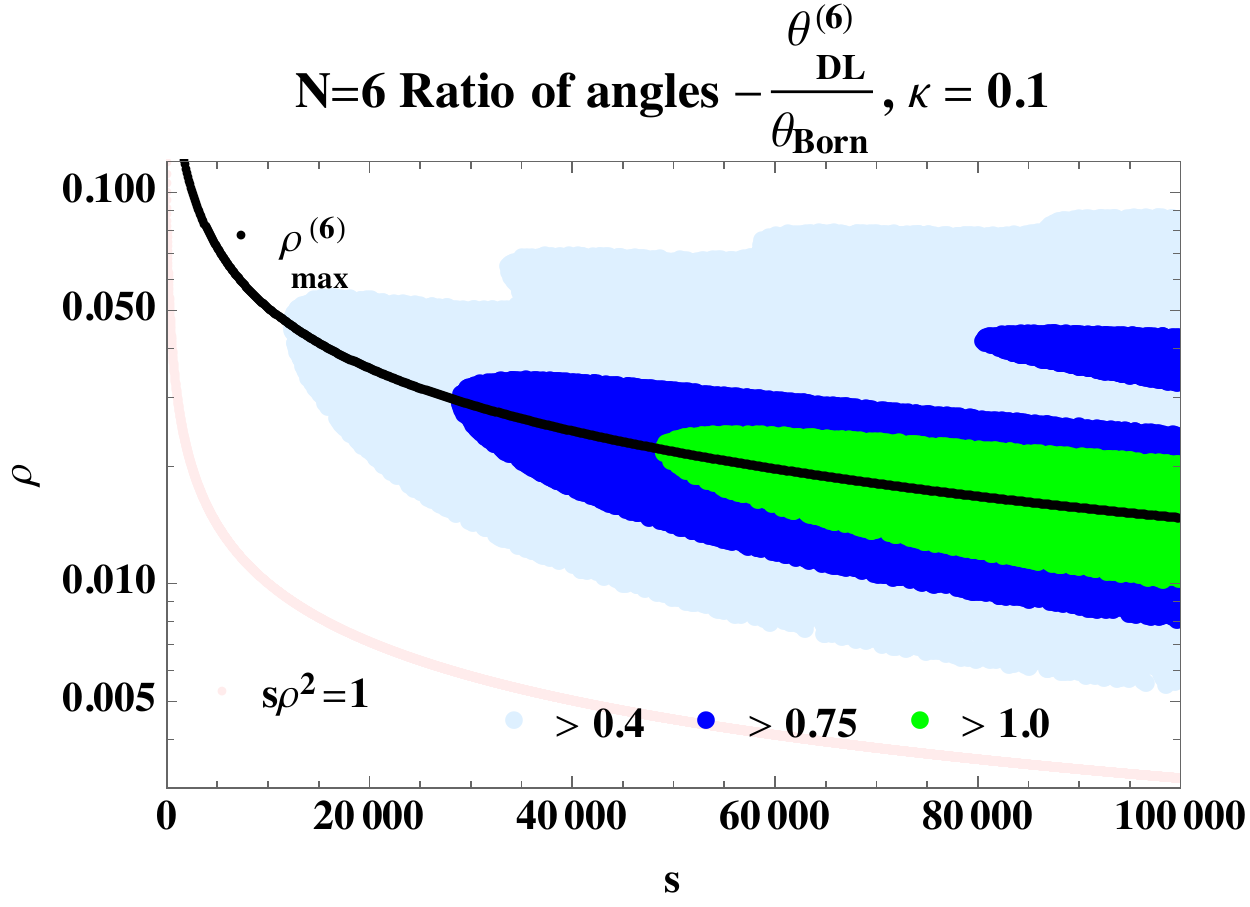}
\end{center}
\vspace{-.5cm}
\caption{$-\frac{\theta^{(N)}_{\rm DL}}{\theta_{\rm Born}}$ as function of $s$ for 
$N$=5 (up) and $N$=6 (bottom) supergravities. }
\label{N56RatioAngles}
\end{figure}

\begin{figure}[h]
\begin{center}
\includegraphics[width=9cm]{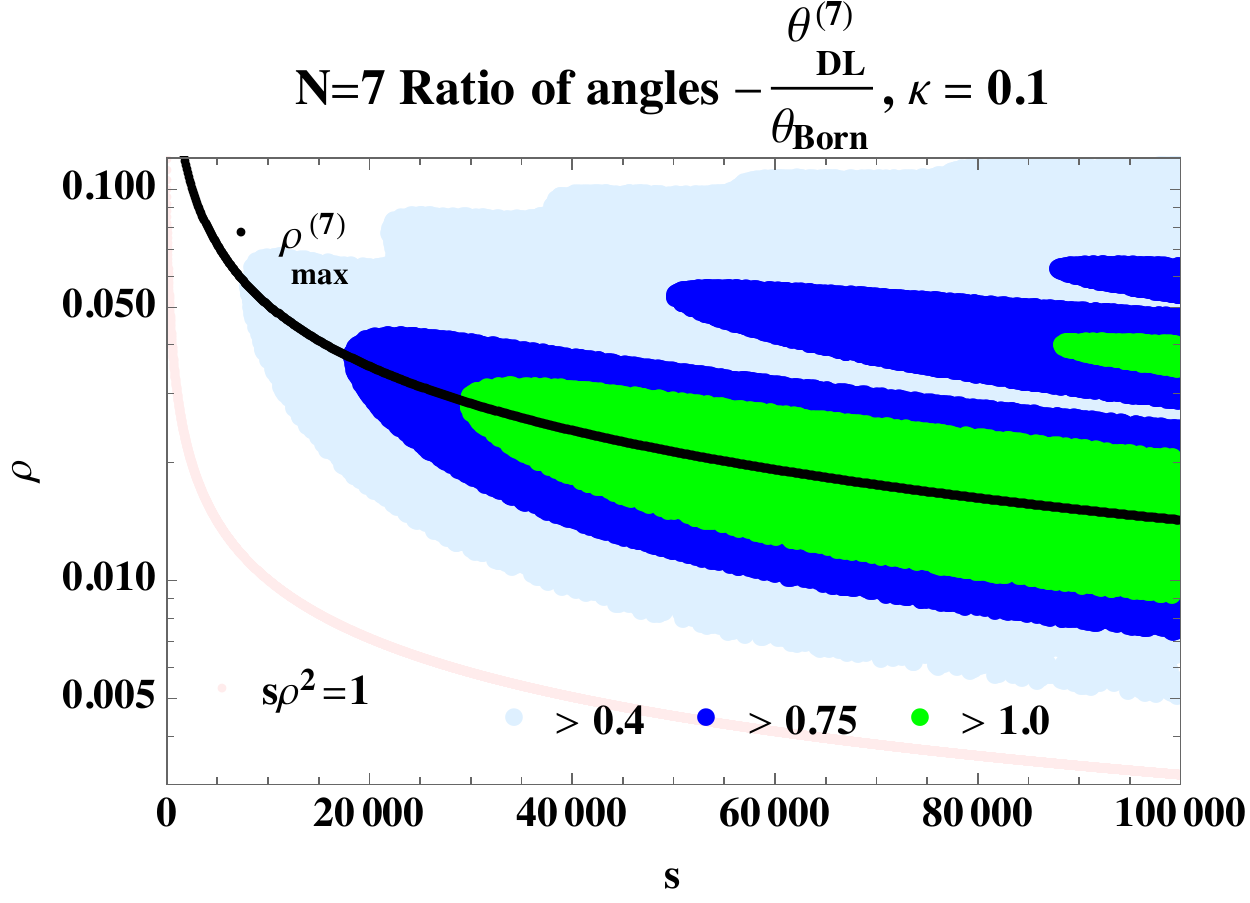}
\includegraphics[width=9cm]{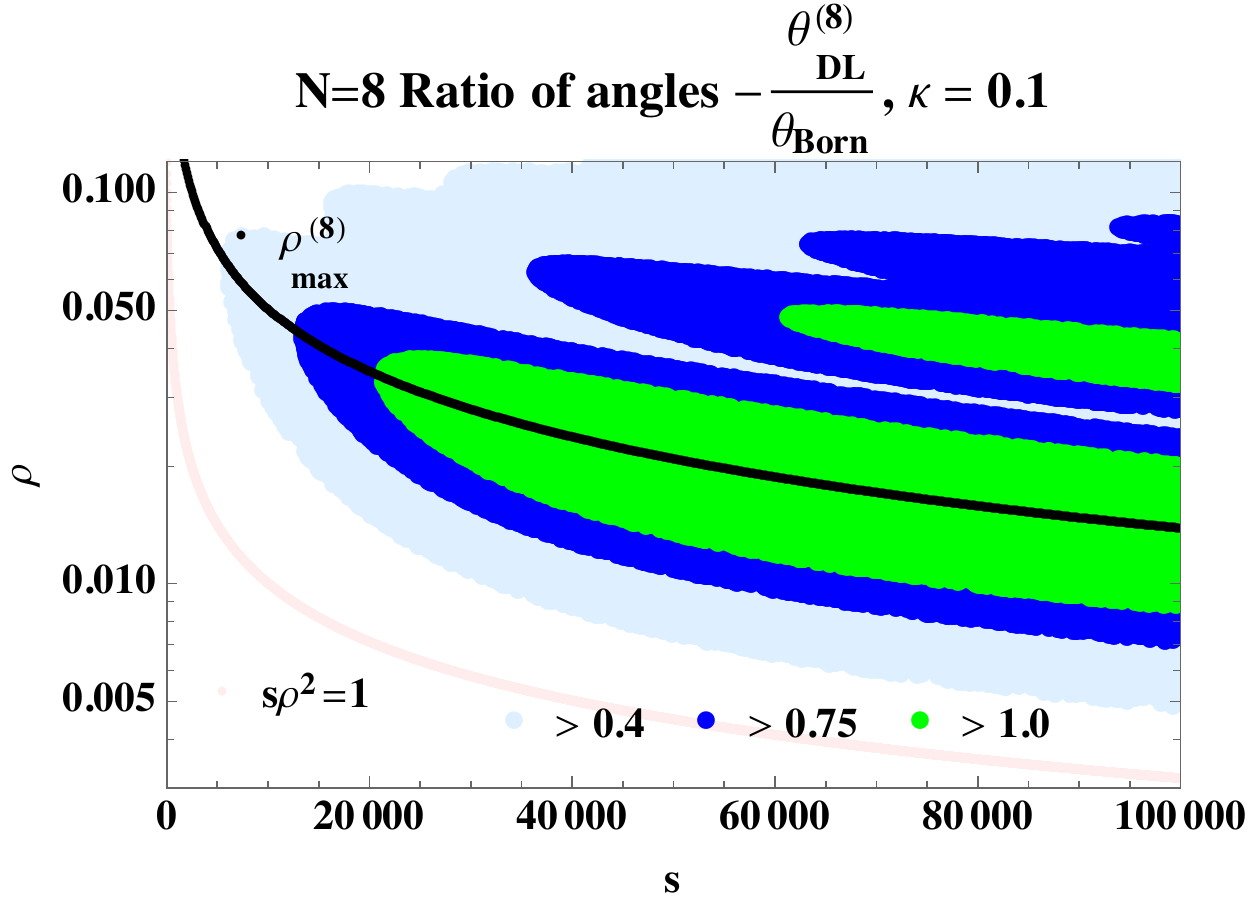}
\end{center}
\vspace{-.5cm}
\caption{$-\frac{\theta^{(N)}_{\rm DL}}{\theta_{\rm Born}}$ as function of $s$ for 
$N$=7 (up) and $N$=8 (bottom) supergravities.}
\label{N78RatioAngles}
\end{figure}

The simplest case to study 
is that of $N=4$ where an infrared scale $\lambda$ is needed to define the phase:\begin{eqnarray}
\chi^{(4)} \left(\rho,s\right) &=& 
 \frac{1 }{2 s}
\int \frac{d^2 \vec{q}}{(2 \pi)^2} e^{-i \vec{q} \cdot \vec{\rho}} 
{\cal A}_4^{\rm Born} (s,t) ~\simeq~
 \frac{\kappa^2 s }{8 \pi^2}
\int \frac{d^2 \vec{q}}{q^2} e^{-i \vec{q} \cdot \vec{\rho}} \, \theta\left(q - \lambda \right)
\nonumber\\
&\simeq& - \frac{\kappa^2 s}{4 \pi} \ln{\left(\rho \, \lambda\right)} 
~=~ \chi_{\rm Born} (\rho,s) \, .
\end{eqnarray}
The graviton's deflection angle in the forward limit is $\lambda$ independent, 
\begin{eqnarray}
\theta^{(4)}(\rho,s)   ~ = ~    \frac{\kappa^2 \sqrt{s}}{2 \pi  \rho}  
~=~ \theta_{\rm Born} (\rho,s) \, .
\end{eqnarray}
Its growth for small $\rho$ is characteristic of an attractive interaction. For general $N$ one can write
 \begin{eqnarray}
\chi^{(N)} \left(\rho,s\right) &=&   \chi_{\rm Born} (\rho,s)  
+ \chi_{\rm DL}^{(N)} \left(\rho,s\right)  \, ,
\end{eqnarray} 
where (see Eq.~(\ref{PertAmplDL4}))
 \begin{eqnarray}
\chi_{\rm DL}^{(N)} \left(\rho,s\right) &=&   \alpha s 
\int \frac{d^2 \vec{q}}{q^2} e^{- i \vec{q} \cdot \vec{\rho}} 
 \sum_{n=1}^\infty {\cal C}_n^{(N)} \, 
\frac{(\alpha \, t)^n}{(2n)!} \ln^{2n}{\left(\frac{s}{-t}\right)} \,  \theta \left(\sqrt{s}-q\right)  
\nonumber\\
 &=&      \pi
\sum_{m=1}^\infty    \sum_{n=0}^{m-1} 
\frac{ (-1)^{m} (\alpha s)^{m+1} {\cal C}_{m-n}^{(N)}}{(n!)^2 m^{2 (m-n) +1}}   
\left({\rho^2 \over 4 \alpha}\right)^{n}  
\label{chidelcoeffs}
\end{eqnarray} 
is infrared finite. For $N>4$ this is a negative function which goes rapidly to zero as $\rho$ grows. The large distance behaviour corresponds to the Born amplitude. This phase  is shown in Fig.~\ref{EikonalPhaseSeveralN} for a large center of mass energy. The DL resummation generates a constant phase shift which is larger as $N$ grows.  
The deflection angle can be written in the form
\begin{eqnarray}
\theta^{(N)} \left(\rho,s\right)  &=& \theta_{\rm Born} \left(\rho,s\right) +
\theta_{\rm DL}^{(N)} \left(\rho,s\right) \, ,\\
\theta_{\rm DL}^{(N)} \left(\rho,s\right) 
&=&   - \theta_{\rm Born} \left(\rho,s\right)  
\sum_{m=2}^\infty    \sum_{n=1}^{m-1} 
\frac{ n  (- \alpha s)^{m} {\cal C}_{m-n}^{(N)}}{(n!)^2 m^{2 (m-n) +1}}    
 \left({\rho^2 \over 4 \alpha}\right)^{n} \, .
 \label{ThetaDLN}
\end{eqnarray} 
\begin{figure}[h]
\begin{center}
\includegraphics[width=9cm]{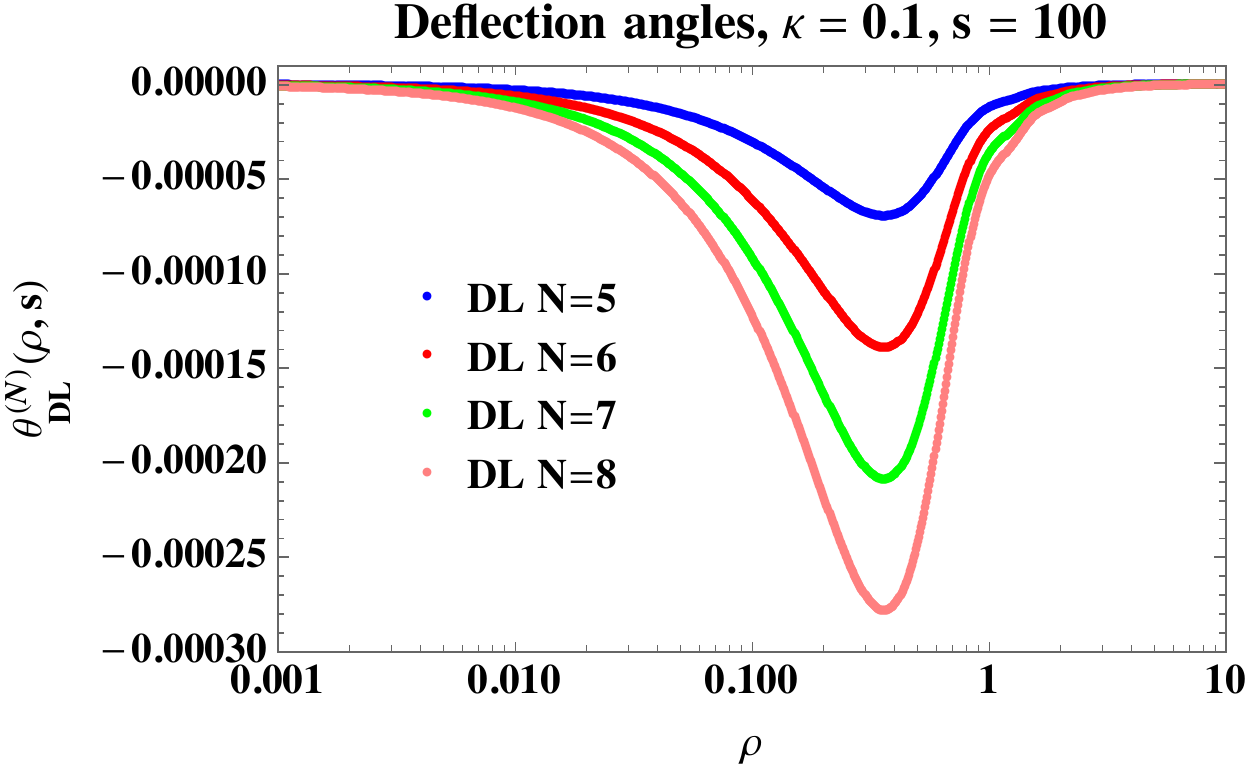}
\end{center}
\vspace{-.5cm}
\caption{DL deflection angle in different supergravities as a function of 
the impact parameter.}
\label{ThetaSeveralN}
\end{figure}
In Fig.~\ref{ThetaSeveralN} it can be seen that $\theta_{\rm DL}^{(N)}$ is negative. 
This means that  the DL resummation has the effect of reducing the Born graviton--graviton deflection angle introducing a screening of the gravitational interaction which is larger for a  larger number of gravitinos. This screening is stronger as $s$ grows and it can make the  graviton's  deflection angle negative. This is shown in Fig.~\ref{DeflectionAngleSeveralN} where $\theta^{(N)}<0$ for some regions in impact parameter space which are larger as $N$ increases. 
The $s$ dependence of the impact parameter $\rho_{\rm max}^{(N)}$ 
where the DLs contribution is maximal is plotted in Figs.~\ref{N56RatioAngles},~\ref{N78RatioAngles}, together with the ratio 
$-\frac{\theta^{(N)}_{\rm DL}}{\theta_{\rm Born}}$ in the $(s,\rho)$ regions where it is large. The DLs make the Born deflection angle to change sign when $-\frac{\theta^{(N)}_{\rm DL}}{\theta_{\rm Born}}>1$. The weaking of gravity is larger in Fig.~\ref{N78RatioAngles}, for $N=7,8$, than in Fig.~\ref{N56RatioAngles}, with $N=5,6$. When $\rho > \rho_{\rm max}^{(N)}$ there are oscillations which appear as bands in $\rho$ space. 

It is important to stress again that the DLs are in principle subleading with respect to the eikonal contributions (see, {\it e.g},~\cite{Ciafaloni:2018uwe,DiVecchia:2019myk,DiVecchia:2019kta}) since they are enhanced by powers of $t$ instead of powers of $s$ (see Eq.~(\ref{Mamplitude})). It is nevertheless an interesting sector of the gravitational scattering amplitudes  which deserves further study since it is a first example of a full resummation of quantum corrections which are infrared finite. It is already remarkable that they reduce the gravitational interaction at high energies in a region of short distances. The fact that this effect is more acute for more gravitinos in the supergravity theory might be related to the possible lack of ultraviolet divergencies in ${\cal N}=8$  supergravity. This is a point which deserves further investigation. 

\begin{figure}[h]
\begin{center}
\includegraphics[width=9cm]{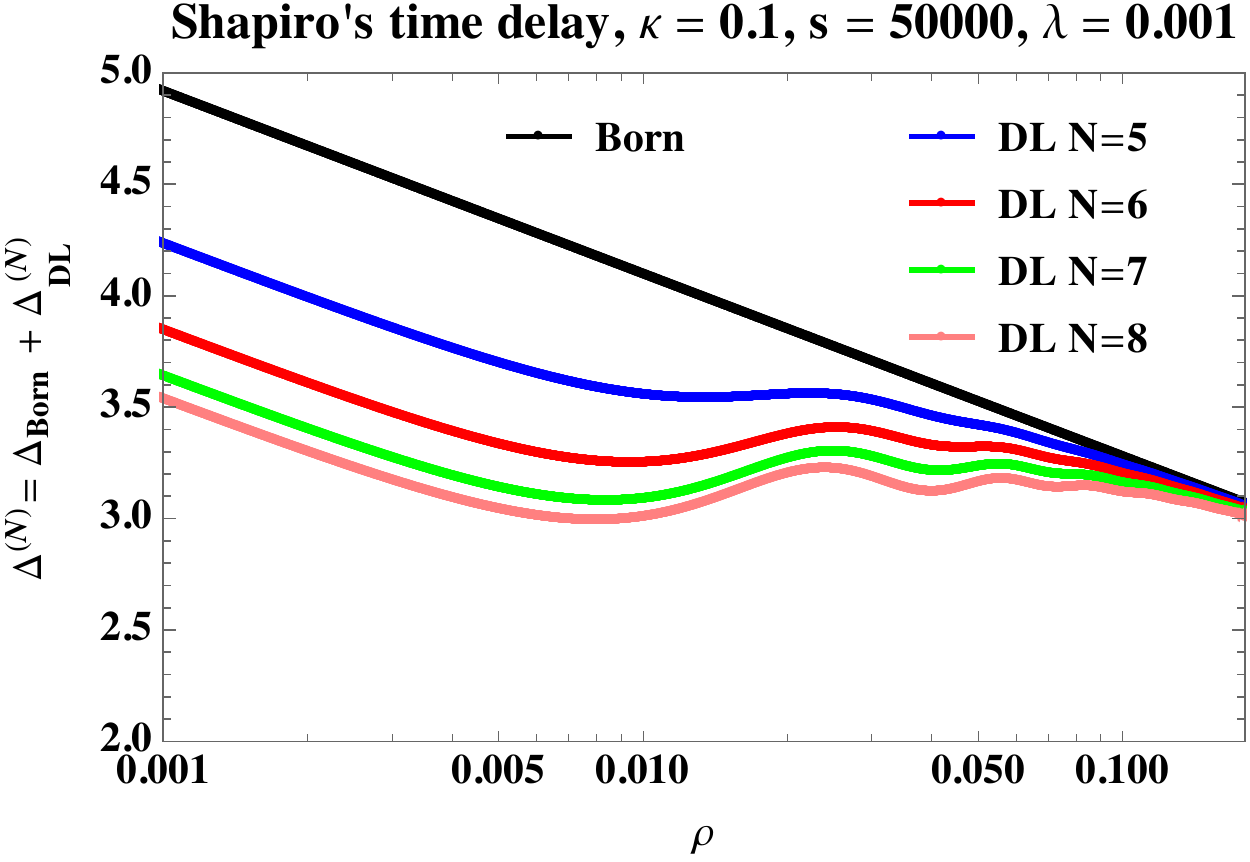}
\end{center}
\vspace{-.5cm}
\caption{Time delay for graviton--graviton scattering for different supergravies with DL accuracy.}
\label{TimeDelaySeveralN}
\end{figure}

A negative contribution to the eikonal phase can be interpreted as a negative contribution to the Shapiro's time delay~\cite{Shapiro:1964uw}  experienced by a particle in elastic scattering~\cite{Kabat:1992tb}.  The explicit formula for the delay experienced by the gravitons at Born level would be
 \begin{eqnarray}
\Delta_{\rm Born} \left(\rho,s\right) &=&  
\frac{\partial }{\partial \sqrt{s}}  \chi_{\rm Born} \left(\rho,s\right)  ~=~ 
- \frac{ \kappa^2 \, \sqrt{s}}{2 \pi} \ln{\left(\rho \, \lambda \right)} \,,
\end{eqnarray}
with DL corrections  of the form
 \begin{eqnarray}
\Delta^{(N)} \left(\rho,s\right) &=&  \frac{\partial }{\partial \sqrt{s}}  \chi^{(N)} \left(\rho,s\right) ~=~   \Delta_{\rm Born} (\rho,s)  + 
 \Delta_{\rm DL}^{(N)} (\rho,s)  \nonumber\\
 &=&     \Delta_{\rm Born} (\rho,s)  +      2  \pi \alpha \sqrt{s} 
\sum_{m=1}^\infty    \sum_{n=0}^{m-1} 
\frac{  (m+1) (-\alpha s)^{m} {\cal C}_{m-n}^{(N)}}{(n!)^2 m^{2 (m-n) +1}}   
\left({\rho^2 \over 4 \alpha}\right)^{n}  \, .
\label{Timedelayone}
\end{eqnarray} 
A numerical representation is shown in  Fig.~\ref{TimeDelaySeveralN}. A constant negative shift appears which at very small $\rho$ becomes 
 \begin{eqnarray}
\Delta^{(N)} \left(\rho,s\right) &\overset{ \rho \ll 1}{\simeq}&  
 \Delta_{\rm Born} (\rho,s)  
+      2  \pi \alpha \sqrt{s} 
\sum_{m=1}^\infty   
\frac{  (m+1) (-\alpha s)^{m} {\cal C}_{m}^{(N)}}{ m^{2 m +1}}  \\
&=&  \frac{\kappa^2\, \sqrt{s}}{4 \pi}\left(
  -  2\ln{\left(\rho \, \lambda \right)}   
+       \int_0^\infty\mu^{(N)} (x) \, dx
\sum_{m=1}^\infty     \left(1+\frac{1}{m}\right) \left(\frac{-\alpha s x}{m^2}\right)^{m}  \right)\, .  \nonumber
\end{eqnarray} 
Quantum corrections contribute to all orders to this expression. In the regions under study the overall time delay is always positive, without challenging the causal structure of the theory (a negative time delay would imply superluminical propagation). 

\section{Connecting with ribbon graphs in ${\cal N}=8$ supergravity}

A surprising correspondence between the evolution equation for $f_\omega^{(N=8)}$ and the theory of graphs embedded on orientable surfaces was found in~\cite{SabioVera:2019edr}.  A map on an orientable surface is a partition defining vertices, edges and faces on it. Edges can split into two half--edges with a fixed orientation, forming a ribbon graph. If an arrow is assigned to a half--edge it is called rooted.  When enumerating equivalence classes of isomorphic rooted maps~\cite{Arques:2000Ar} a relation equivalent to Eq.~(\ref{evoleq}) appears. 

A graphical bijection between the DL Feynman diagrams and ribbon graphs can be drawn. Similar mappings take place in other theories~\cite{Prunotto:2013tpa,Gopala:2017mjh,GopalaKrishna:2018wzq}. The Born diagram maps to the one root ribbon graph with zero edges,
  \begin{eqnarray}
 \hspace{-.cm}\parbox{15mm}{\includegraphics[width=2.2cm,angle=0]{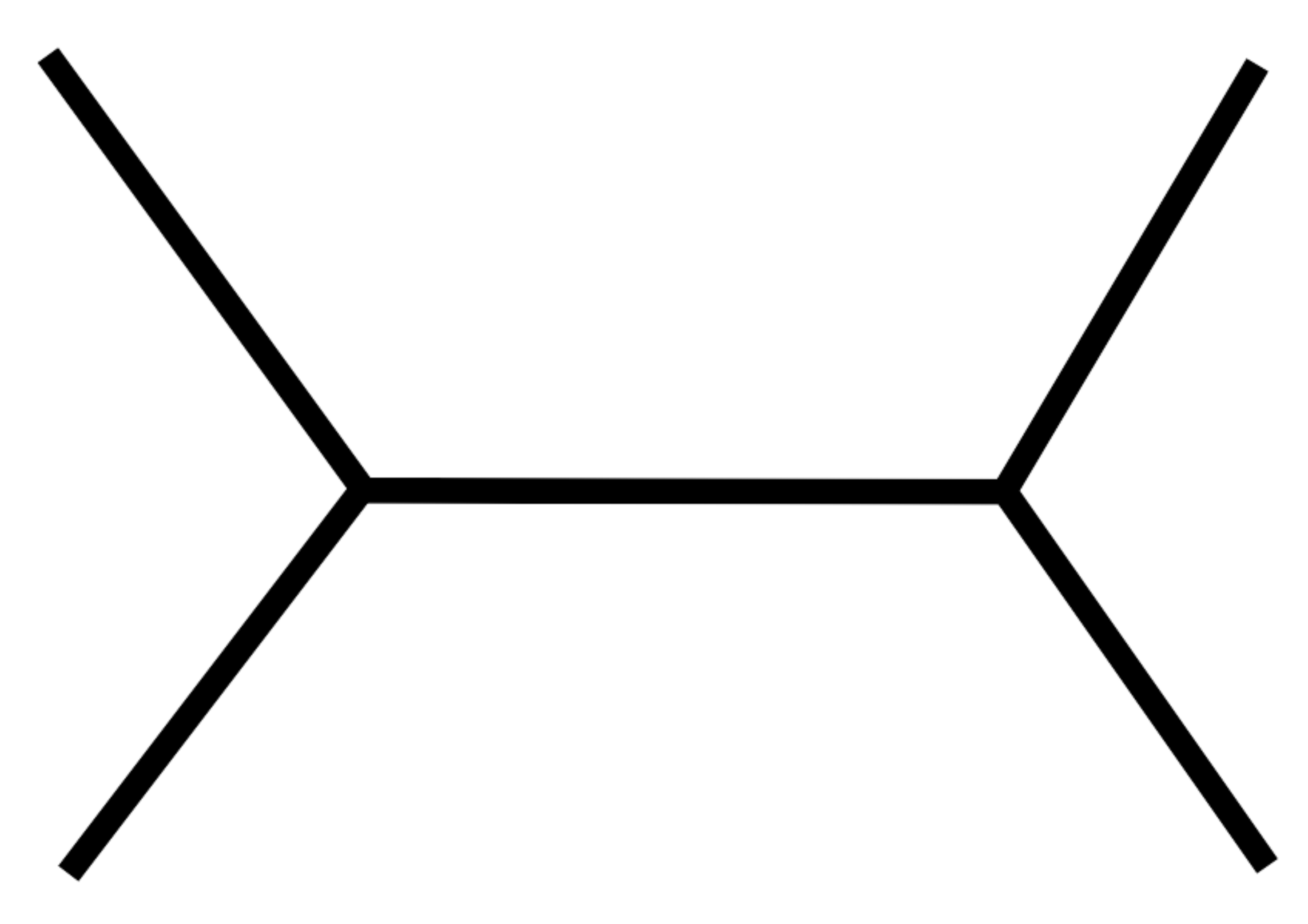}} \hspace{.8cm}
 &=& \hspace{.3cm} \parbox{15mm}{\includegraphics[width=.6cm,angle=0]{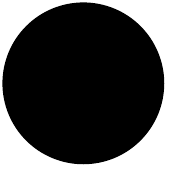}} 
\end{eqnarray}
A first set of DLs is generated by attaching soft gravitons and gravitinos to the external lines. The sum of the associated Feynman diagrams is then mapped to the one--edge graph with one vertex,
\begin{eqnarray}
   2 \hspace{.cm} \parbox{15mm}{\includegraphics[width=2.2cm,angle=0]{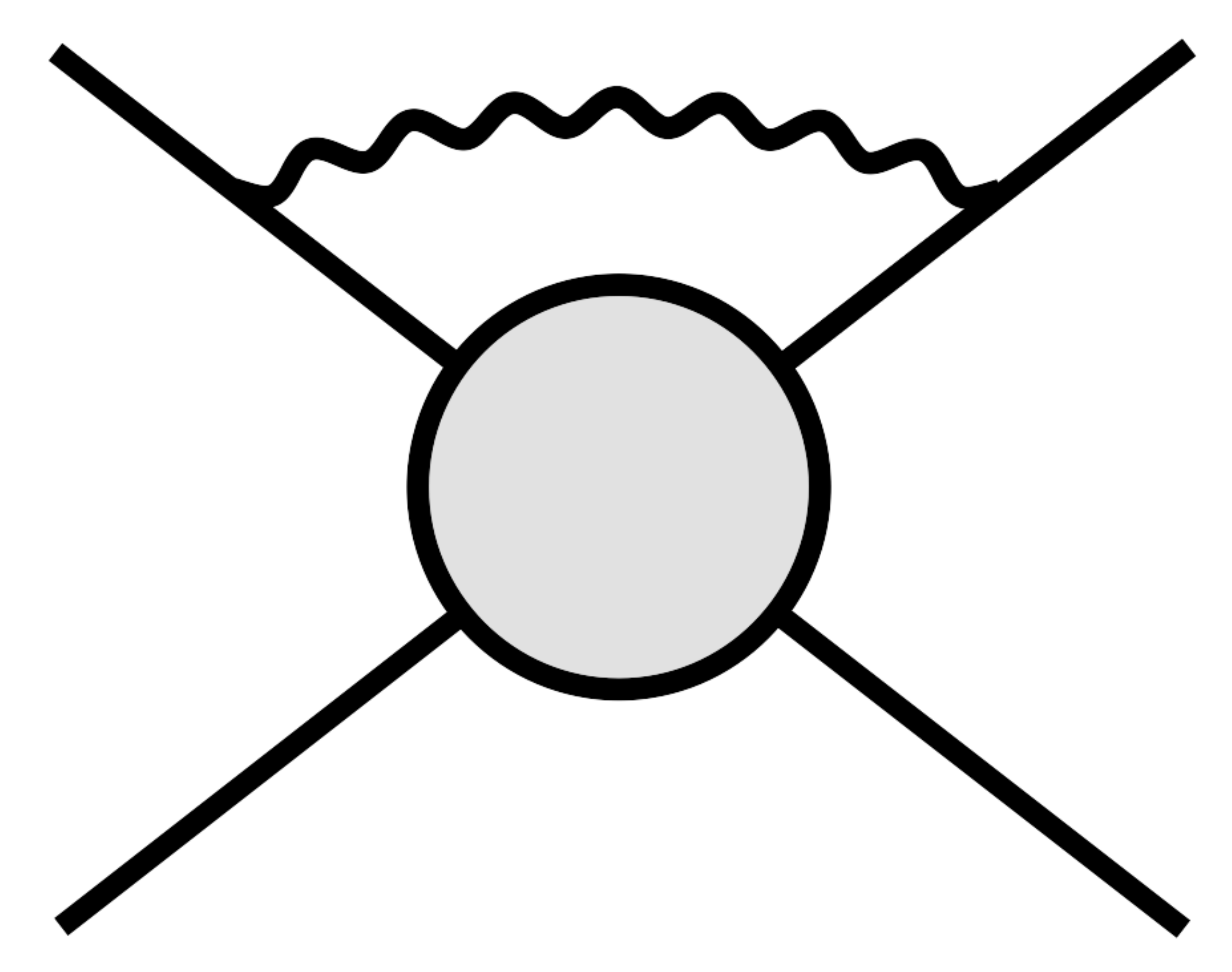}}
 \hspace{0.5cm}
 + 2 \hspace{.cm} \parbox{15mm}{\includegraphics[width=2.2cm,angle=0]{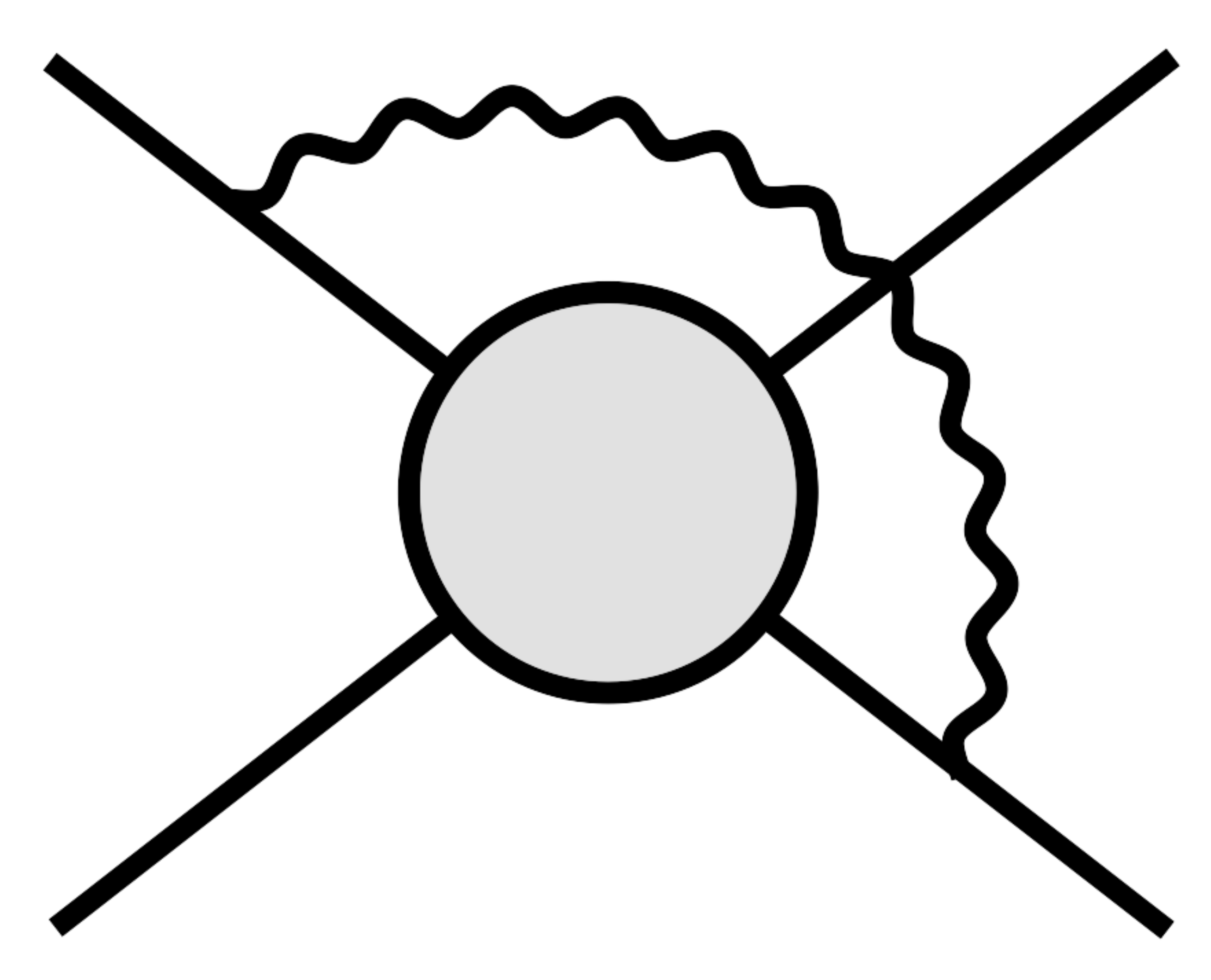}}
 \hspace{0.8cm}
  &=& \hspace{.3cm} \parbox{15mm}{\includegraphics[width=1.4cm,angle=0]{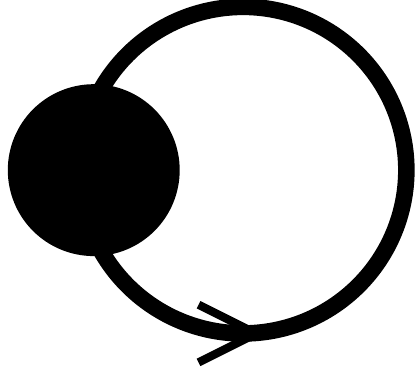}}\hspace{1cm}\hspace{-.8cm}
\end{eqnarray}
A second set of DLs stems from the exchange in the $t$--channel of two gravitons or gravitinos. The corresponding Feynman diagram is mapped to the two--vertex 1--rooted ribbon graph:
  \begin{eqnarray}
 \hspace{-.cm}\parbox{15mm}{\includegraphics[width=2.8cm,angle=0]{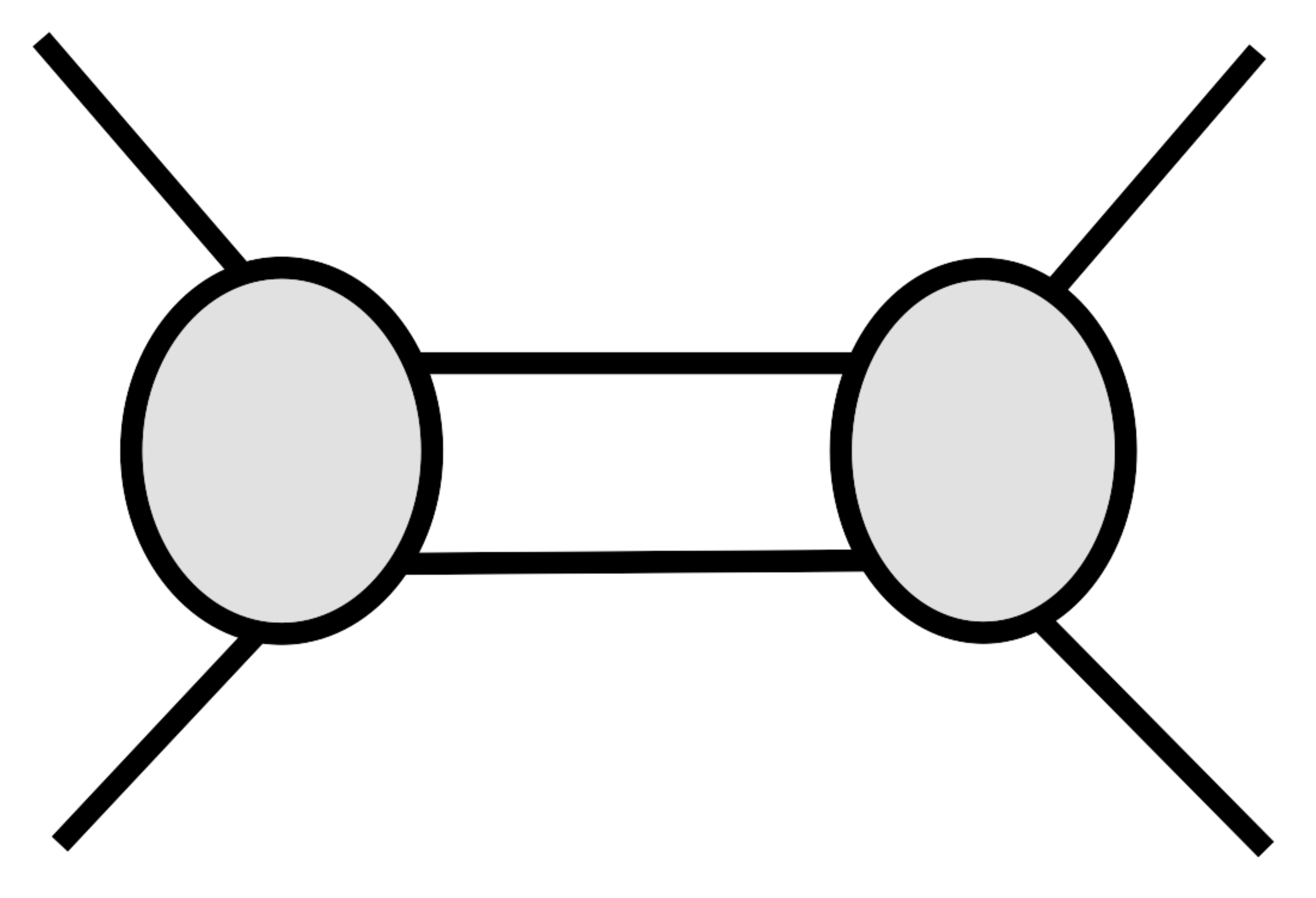}} \hspace{1.4cm}
 &=& \hspace{.3cm} \parbox{15mm}{\includegraphics[width=2.8cm,angle=0]{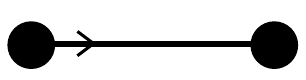}} 
\end{eqnarray}
These two configurations generate the second term in the series $(1,2,10,74,\dots)$ of Eq.~(\ref{fomegaexp}). Further coefficients appear by iteration of these two basic elements. For example, the coefficient 10 stems from
  \begin{eqnarray}
 \hspace{-3.cm}\parbox{30mm}{\includegraphics[width=4.2cm,angle=0]{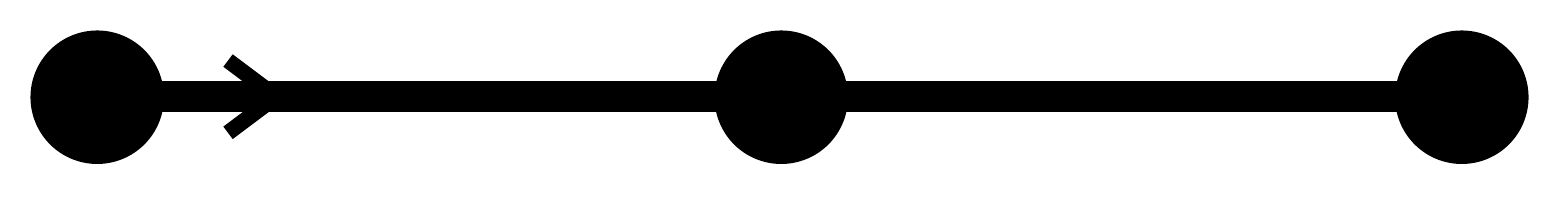}} \hspace{1.4cm}
 && \hspace{.3cm} \parbox{30mm}{\includegraphics[width=4.2cm,angle=0]{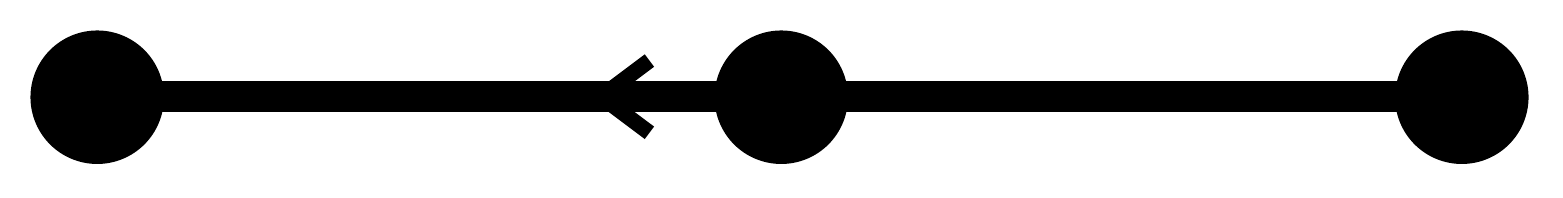}} \nonumber\\
 \hspace{.4cm}\parbox{25mm}{\includegraphics[width=3.1cm,angle=0]{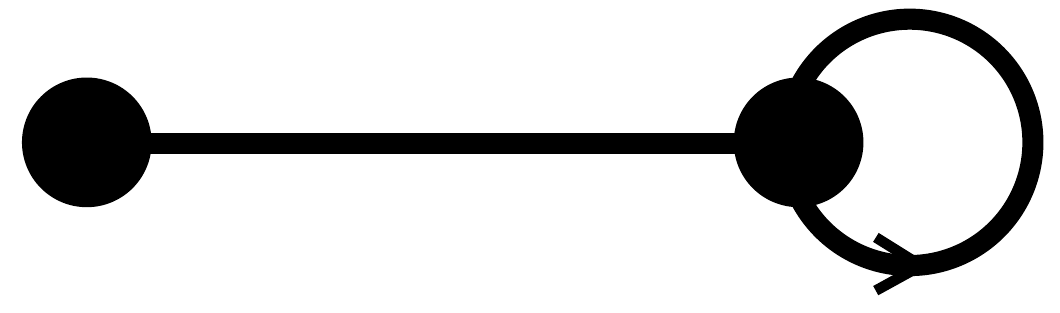}} \hspace{1.4cm}
 && \hspace{.8cm} \parbox{25mm}{\includegraphics[width=3.1cm,angle=0]{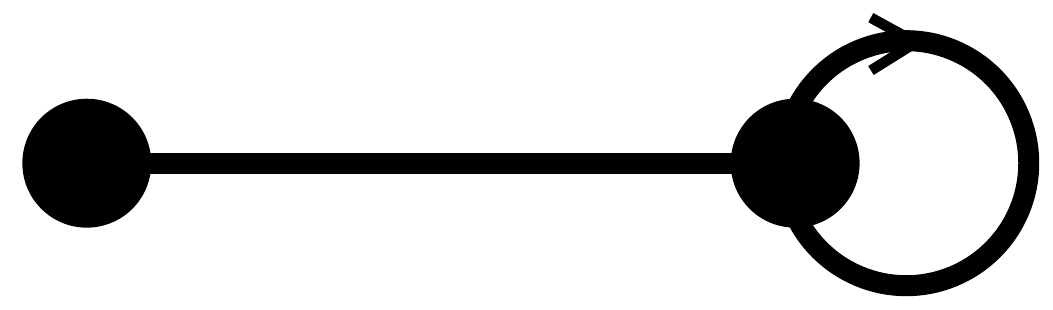}} \nonumber\\
  \hspace{-.cm}\parbox{25mm}{\includegraphics[width=3.1cm,angle=0]{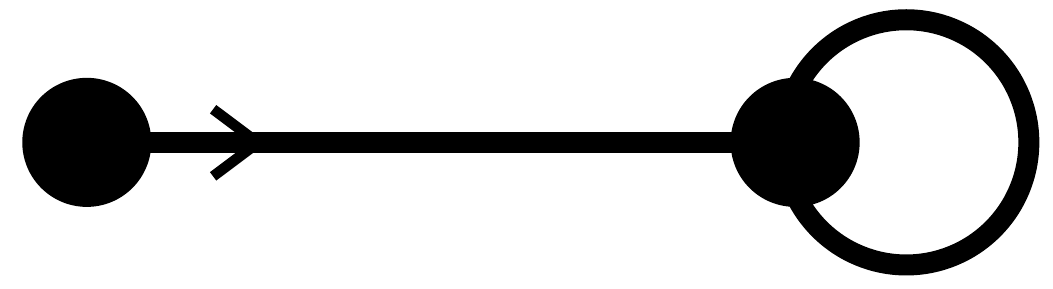}} \hspace{1.4cm}
 && \hspace{.8cm} \parbox{25mm}{\includegraphics[width=3.1cm,angle=0]{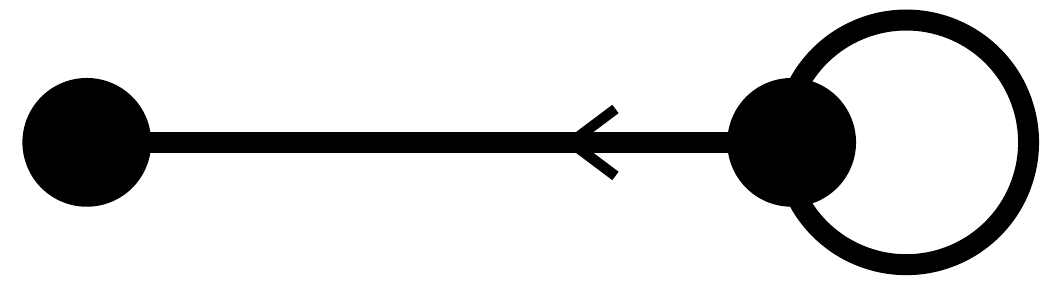}} \nonumber\\
  \hspace{-.cm}\parbox{19mm}{\includegraphics[width=1.8cm,angle=0]{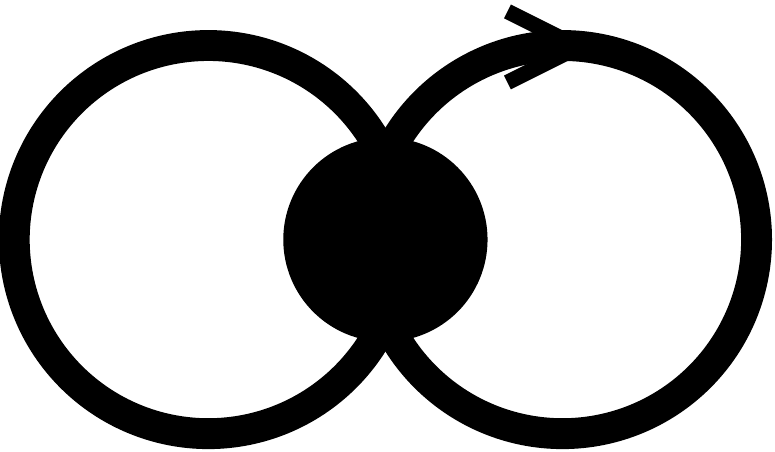}} \hspace{1.4cm}
 && \hspace{1.4cm} \parbox{19mm}{\includegraphics[width=1.8cm,angle=0]{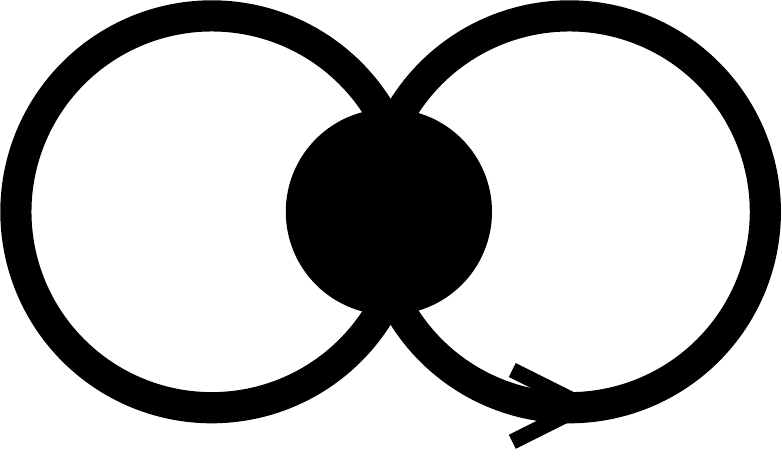}} \nonumber\\
  \hspace{-.2cm}\parbox{16.mm}{\includegraphics[width=1.3cm,angle=0]{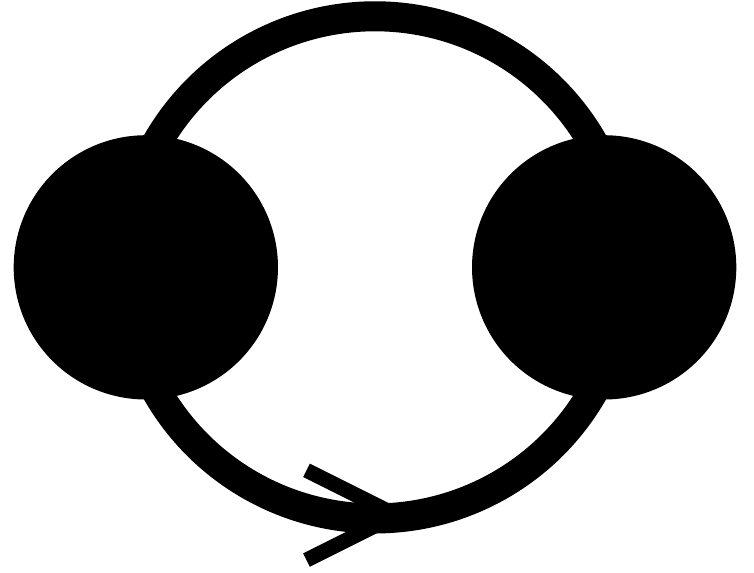}} \hspace{1.4cm}
 && \hspace{1.5cm} \parbox{15mm}{\includegraphics[width=1.6cm,angle=0]{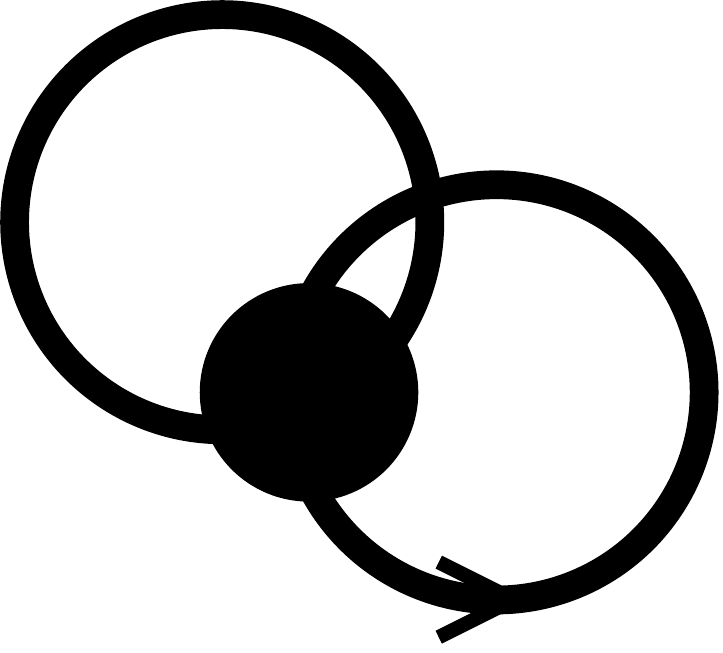}} \nonumber
\end{eqnarray}
This representation is simpler than the Feynman diagram approach and it will be worth studying further its implications.  

\section{The last scientific discussion with Lev}

At the beginning of August 2017 I had my last scientific discussion with Lev. It was at DAMTP, in Cambridge. He had just delivered a seminar on the high energy effective action for gravity and we had a kind of brainstorming session on the blackboard while Elvira, Lev's wife, was making sure all the details of their bus and plane tickets were correct. She always was Lev's connection to planet Earth. 

A couple of days before we had been in Southampton visiting Douglas Ross and his wife Jackie and discussing about the meaning of introducing temperature in the high energy limit of Quantum Chromodynamics. That was one of Lev's favourite subjects. He was proud of his work with Hector de Vega where they were able to introduce temperature in the $t$--channel of a scattering process without breaking the integrability of the BFKL equation. Our discussions with Doug centered around the meaning of this temperature as a parameter to introduce confinement in the whole framework. 

During his talk Lev was asked many good questions mainly by Michael Green and Hugh Osborn, who could appreciate the huge amount of work he had devoted to the subject and the implications of his results (a true {\it tour de force} in Green's words).
 
Lev was interested in all new developments and wanted to know about the  attempts to connect the high energy effective action with the so--called CHY formalism. In the following the first baby steps in this work~\cite{Chachamis:2017jbn}, which were published with Grigorios Chachamis, Miguel {\'A}ngel V{\'a}zquez--Mozo (both great physicists with whom it is a pleasure to discus anything, not only science) and Diego Medrano (doctoral student at the time and now fighting in the postdoc jungle) will be briefly described. But before doing so let me  write that Lev was very enthusiastic about the possibilities of this line of thinking and we were planning on his next visit to Madrid to work together on this. Needless to write that to make progress in this or any other direction will be much harder without Lev's wisdom and mastery to guide us. 

They almost missed the bus to the airport, we had to run along Parker's  Piece with all the luggage to catch it. It was there, completely breathless, where we shook hands for the last time. 

\subsection{CHY Scattering Equations in Sudakov variables}

The standard Feynman diagram techniques for the evaluation of scattering amplitudes become too complicated when the number of legs or loops grows. The Cachazo--He--Yuan (CHY) approach~\cite{Cachazo:2013hca,Cachazo:2013iea}  
offers a promising route to simplify this problem where $n$--point amplitudes are presented as $(n-3)$--dimensional integrals over the moduli 
space of $n$--punctured spheres, which are localized on the solutions to the so-called {\sl scattering equations} (SE). 

The bottleneck in this framework is that the number of integrals defining the $n$--point $S$-matrix elements grows very rapidly with $n$. The value of~\cite{Chachamis:2017jbn} is to propose the use of Sudakov variables to simplify the problem of finding solutions of the SE since they naturally encode momentum conservation. The SL(2,${\mathbb C}$) redundancy of the SE is only partially fixed in order to use azimuthal angles (with respect to the axis defined by the incoming particles) to parametrize the position of the punctures on the sphere.

For CHY the set of $n$ on--shell $D$--dimensional momenta $p_i^{\mu}$ can be mapped into an internal space on the $n$--punctured Riemann sphere parametrized by $\sigma_i \in \mathbb{C}\mathbb{P}^1$ (with $i=1,\ldots,n$) using 
\begin{eqnarray}
p^\mu_j  = \oint\limits_{\left|z-\sigma_j \right|=\epsilon} \frac{dz}{2 \pi i} \frac{v^\mu(z)}{\prod_{k=1}^n (z-\sigma_k)} \, \, \, , \, \, \,
v^\mu(z) = \sum_{j=1}^{n}p_j^\mu \prod_{\substack{k=1\\k\neq j}}^n (z-\sigma_k),
\end{eqnarray} 
where $v(z)^{2}=0$. This implies the SE ($\sigma_{ij} \equiv \sigma_i - \sigma_j$, $s_{ij}=(p_{i}+p_{j})^{2}=2p_{i}\cdot p_{j}$)
\begin{align}
{\cal S}_i (\sigma) \equiv \sum_{j \neq i}^n \frac{s_{ij}}{\sigma_{ij}}=0 \, .
\label{eq:S_i=0}
\end{align}
There are $n$ equations with $n$--3 being linearly independent. They have $(n-3)!$ solutions. 

CHY shown~\cite{Cachazo:2013hca,Cachazo:2013iea} that the tree-level $n$-point Yang-Mills amplitudes admit the following representation with support on the SE solutions: 
\begin{align}
{\cal A}_n = i \, g^{n-2} \int \frac{d^{n} \sigma}{{\rm Vol} [{\rm SL}(2,\mathbb{C})]}
\sigma_{kl} \sigma_{lm} \sigma_{mk} {\prod_{i \neq k,l,m}}
\delta \left(\sum_{j \neq i}^n \frac{2 \, p_i \cdot p_j}{\sigma_{ij}}\right) I_L I_R \, ,
\label{eq:general_amplitude_SE}
\end{align}
where 
\begin{eqnarray}
I_L = \sum_{\beta \in S_n/\mathbb{Z}_n} \frac{{\rm Tr} \left(
T^{a_{\beta(1)}} T^{a_{\beta(2)}} \cdots T^{a_{\beta(n)}}\right)}{\sigma_{\beta(1) \beta(2)} \sigma_{\beta(2) \beta(3)} \cdots \sigma_{\beta(n) \beta(1)}} \, \, \, , 
\, \, \,  I_R = {\rm Pf}' M_n \, .
\label{eq:IR_redpfaffian}
\end{eqnarray}
The latter depends on the reduced Pfaffian of a $2n\times 2n$ antisymmetric matrix which is a function of the momenta and polarizations of external particles. 
It is remarkable that for amplitudes in Einstein-Hilbert gravity it is enough to make the replacement
\begin{eqnarray}
I_L = {\rm Pf}' M_n \, \, \, , 
\, \, \, 
I_R = {\rm Pf}' M_n \, ,
\end{eqnarray}
exhibiting the double-copy structure of graviton amplitudes. 

In the 4--dimensional case there exists an interesting solution to the SE found by Fairlie~\cite{Fairlie:2008dg,Fairlie:1972zz}. It reads
\begin{align}
\sigma_j &= \frac{p_j^0 + p_j^3}{p_j^1 - i p_j^2}  =  \frac{p_j^1 + i p_j^2}{p_j^0 - p_j^3} \, .
\label{eq:sln_fairlie}
\end{align}
Using the parameterization of on-shell momenta $p_j$ 
\begin{align}
p_j = p_{j}^{\perp} \left(\cosh{Y_j},\cos{\phi_j},\sin{\phi_j},\sinh{Y_j}\right) \, ,
\label{eq:ps_rapidity_azimuthal_def}
\end{align}
with $Y_j$ the rapidity, $\phi_{j}$ the azimuthal angle, $p_{j}^{\perp}$ the modulus of the transverse component of the momentum, and the stereographic coordinates on $\mathbb{S}^{2}$,
\begin{align}
p_j= \omega_j \left(1, \frac{\zeta_j + \bar{\zeta}_j}{1+ \zeta_j \bar{\zeta}_j},i \frac{ \bar{\zeta}_j-\zeta_j}{1+ \zeta_j \bar{\zeta}_j},\frac{\zeta_j \bar{\zeta}_j-1}{1 + \zeta_j \bar{\zeta}_j}\right) \, ,
\end{align}
one can write Fairlie's solution~\eqref{eq:sln_fairlie} in the form  
\begin{align}
\sigma_j = \zeta_j = e^{Y_j + i \phi_j} \, .
\label{eq:punctures_position_Yphi}
\end{align}
$p_j$ is then mapped onto the point $2\sigma_{j}$ on the projective  plane. On the sphere it lies on a circumference of radius $2e^{Y_{j}}$ parametrized by the azimuthal angle $\phi_{j}$. 

The two incoming particles with momenta $p$ and $q$ in a general process 
in which the particles in the final state have momenta $p_{i}$ (with $i=1,\ldots,n-2$) are considered to have their spatial momenta lying along the $z$ axis taking the form
\begin{eqnarray}
p \longrightarrow  \frac{\sqrt{s}}{2} (1,0,0,1)
\, \, \, , \, \, \, 
q \longrightarrow \frac{\sqrt{s}}{2} (1,0,0,-1).
\label{eq:incoming_momenta}
\end{eqnarray}
The associated punctures solving the SE are located on a circle around the north and south poles of the Riemann sphere, which shrinks to a point when $\epsilon\rightarrow 0$, {\it i.e.}
\begin{eqnarray}
\sigma_p = e^{Y_p + i \phi} = \frac{e^{i \phi}}{\epsilon} \longrightarrow \infty
\, \, \, , \, \, \, 
\sigma_q = - e^{Y_q + i \phi}=  - \epsilon \, e^{i \phi} \longrightarrow 0.
\end{eqnarray}
In a general 4--point scattering amplitude the incoming and outgoing momenta given by $\{p, q, p', q'\}$ and constrained by $p + q - p' - q' =0$ are mapped into the moduli space of spheres with four punctures  $\{\sigma_p, \sigma_q,\sigma_{p'} ,\sigma_{q'}\} \in \mathbb{C}\mathbb{P}^1$. 
The incoming momenta can be treated as above and the outgoing particles using  the Sudakov~\cite{Sudakov:1954sw} representation as follows,
\begin{eqnarray}
q_1 &\equiv& p - p' ~=~ \alpha \,  p + \beta \, q + \mathbf{q}_1
\, \, \, , \, \, \, 
\mathbf{q}_1 ~=~ q_1^\perp \left(0,\cos{\theta_1},\sin{\theta_1},0\right) \, , \nonumber\\
p' &=& p - q_1  ~=~  \left({\sqrt{s}\over 2}\right.\left.(1-\alpha-\beta),-q_{1}^{\perp}\cos\theta_{1},-q_{1}^{\perp}\sin\theta_{1},
{\sqrt{s}\over 2}(1-\alpha+\beta)\right) \, . 
\nonumber
\end{eqnarray}
The on-shell condition leads to $|Q_{1}|^{2}\equiv (q_{1}^{\perp})^{2}=s(\alpha-1)\beta$ with $ Q_{j}=q_j^\perp e^{i \theta_j}$. A similar parameterisation applies for $q' = q + q_1$ with $|Q_{1}|^{2}\equiv (q_{1}^{\perp})^{2}=s\alpha(1+\beta)$. Therefore $\alpha+\beta=0$. 

\begin{figure}
\begin{center}
\vspace{-.3cm}
\includegraphics[scale=.6]{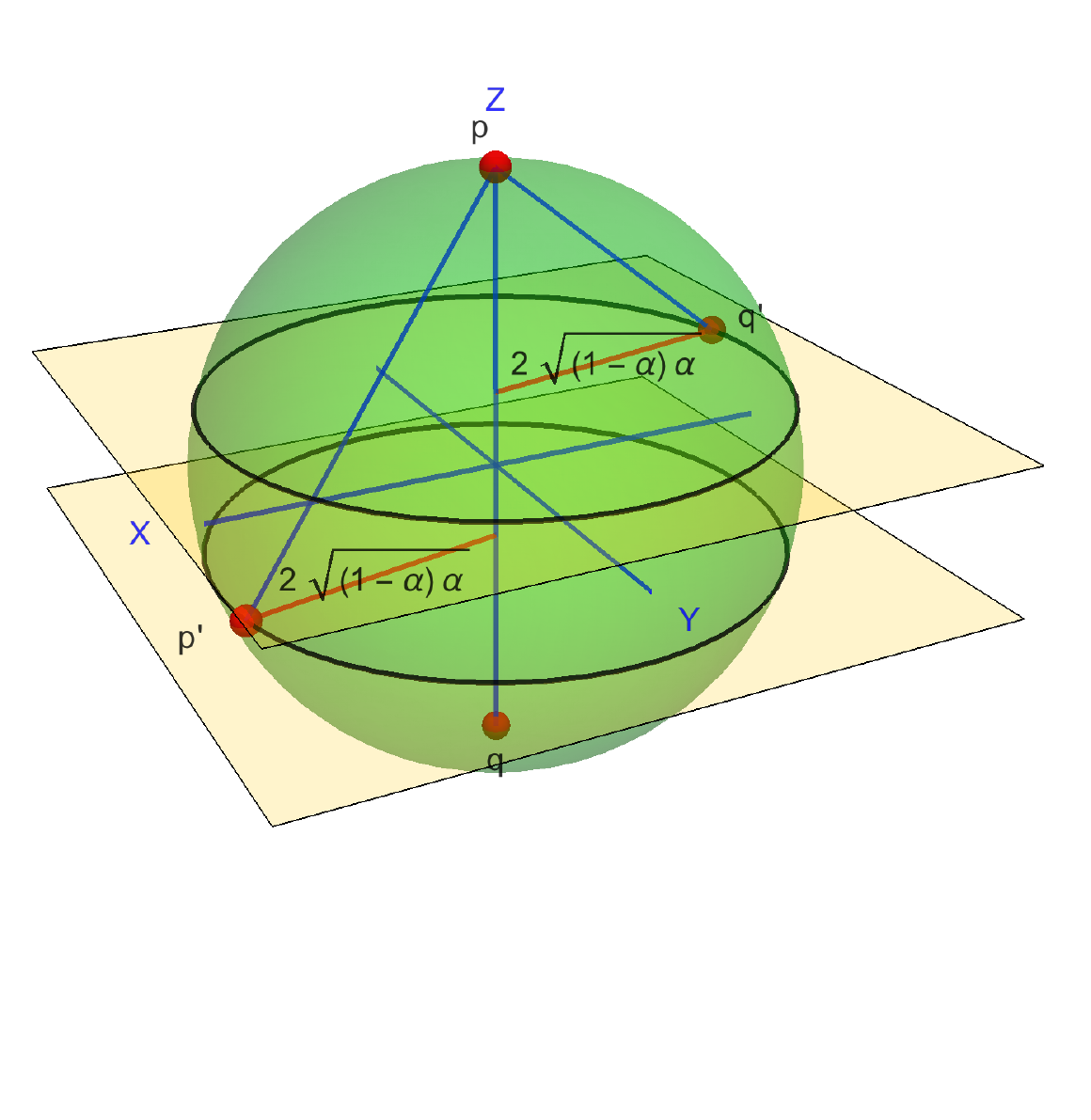}
\vspace{-2.cm}
\caption{Punctures on the Riemann sphere for 4--particle scattering with momenta $p + q \longrightarrow p' + q'$.}
\label{Stereo3}
\end{center}
\end{figure}

For the 4-point function, the SE only have one solution, which corresponds to 
Fairlie's~\eqref{eq:sln_fairlie}: 
\begin{align}
\sigma_{p'} &\equiv e^{Y_{p'} + i \phi_{p'}}
=\frac{ Q_1}{ \beta \sqrt{s}}= \sqrt{\frac{1-\alpha}{\alpha}} e^{i(\theta_1+\pi)} \, ,  \nonumber\\
\sigma_{q'} &\equiv 
 e^{Y_{q'}+ i \phi_{q'}}=  
\frac{ Q_1}{(1-\alpha) \sqrt{s}} = \sqrt{\frac{\alpha}{1-\alpha}} e^{i\theta_1} \, ,
\end{align}
and the two punctures are located on antipodal points on the sphere as can be seen in Fig.~\ref{Stereo3}. 
Using $s_{ij} =s\widehat{s}_{ij}, Q_{i} = \sqrt{s}\widehat{Q}_{i}$, the SE solution now reads
\begin{eqnarray}
\sigma_{p'} = -{\widehat{Q}_{1}\over \alpha}
\, \, \, , \, \, \,
\sigma_{q'} = {\widehat{Q}_{1}\over 1-\alpha} \, .
\end{eqnarray}
In the Sudakov representation the evaluation of scattering amplitudes within the CHY formalism is rather simple. As an example let us consider  the 4--point amplitude in a $\varphi^{3}$ scalar theory:
 \begin{align}
\mathcal{A}_4^{\varphi^3} &=  \int  
d z_{p'}\left[z_{p'}-{Q_{1}\over (1-\alpha)\sqrt{s}}\right]^{-2} \frac{Q_1^2}{s^2 \alpha^3  (\alpha-1)}
\delta \left(z_{p'} + \frac{Q_1}{\alpha \sqrt{s}}\right)
 = 
  \frac{ \alpha-1 }{s \alpha} \, .
 \end{align} 
Notice that the phase in $Q_1$  cancels out in the amplitude.

\begin{figure}[t]
\begin{center}
\vspace{-.3cm}
\includegraphics[scale=.45]{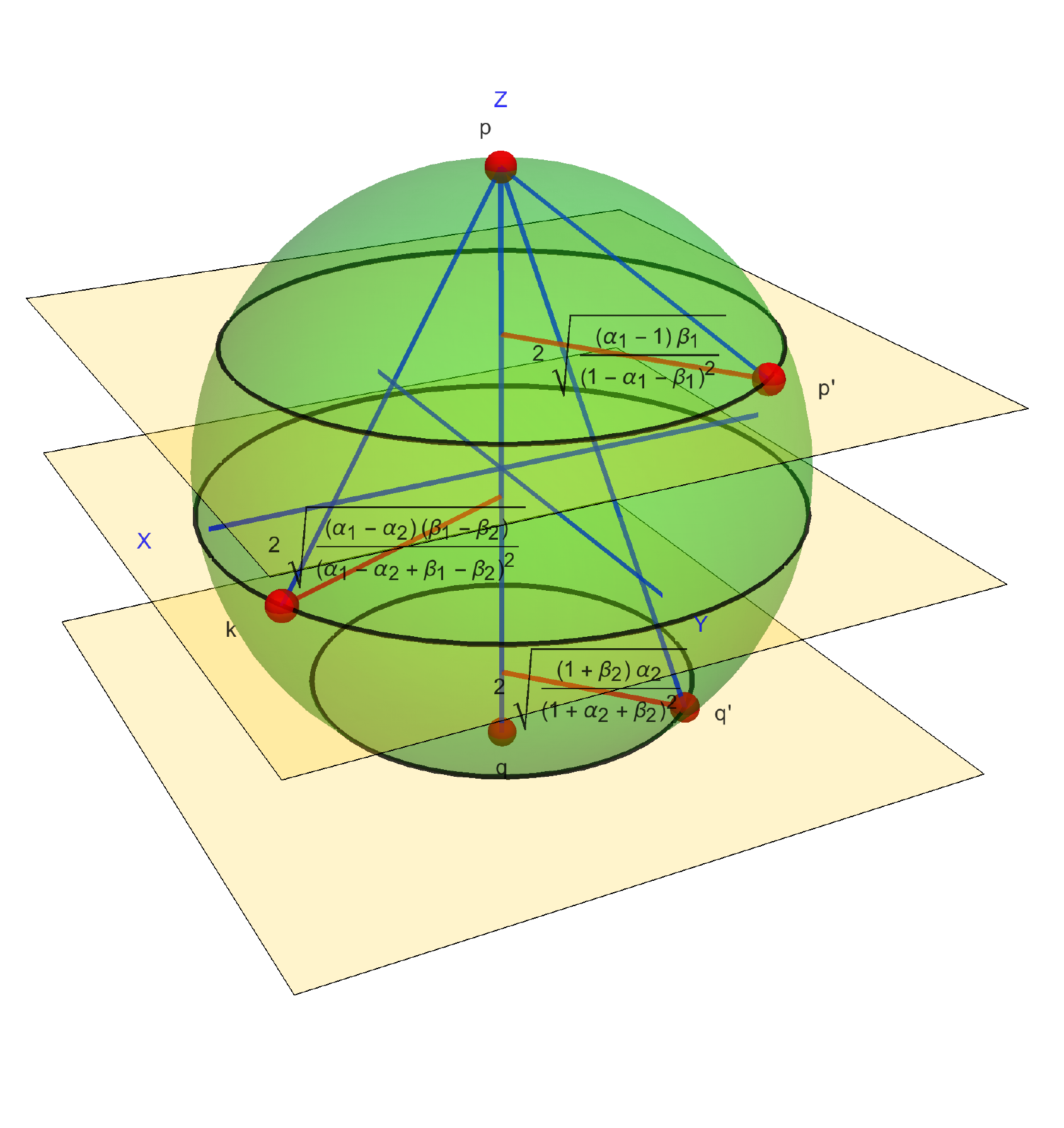}
\vspace{-1.cm}
\caption{Punctures on the Riemann sphere for the 5--particle amplitude.}
\label{Stereo4}
\end{center}
\end{figure}

In the case of five particles, $\{p, q, p', k, q'\}$ with $p+q-p'-k-q'=0$, there exists a second solution to the SE besides Fairlie's. Two sets of Sudakov parameters $\{\alpha_{1},\beta_{1}\}$ and $\{\alpha_{2},\beta_{2}\}$ are introduced, {\it i.e.}
\begin{eqnarray}
q_1 &=& p - p' = \alpha_1 p + \beta_1 q + \mathbf{q}_{1} \, , \nonumber \\
q_2 &=& q' - q = \alpha_2 p + \beta_2 q + \mathbf{q}_{2} \, ,\\
k &=& q_1 - q_2 = \left(\alpha_1 - \alpha_2\right) p + \left(\beta_1 - \beta_2\right) q + \mathbf{q}_1 - \mathbf{q}_{2} \, , \nonumber
\end{eqnarray}
where $\mathbf{q}_i =q_i^\perp \Big(0,\cos{\theta_i},\sin{\theta_i},0\Big)$. The on--shell conditions amount to $|Q_{1}|^{2}=s(\alpha_{1}-1)\beta_{1}$, 
$|Q_{2}|^{2}=s\alpha_{2}(1+\beta_{2})$, $|Q_{1}-Q_{2}|^{2}=s(\alpha_{1}-\alpha_{2})(\beta_{1}-\beta_{2})$ and $Q_{1}Q_{2}^{*}+Q_{1}^{*}Q_{2}= s(\alpha_{2}-\beta_{1}+\alpha_{1}\beta_{2}+\alpha_{2}\beta_{1})$. The punctures associated to Fairlie's solution are then
\begin{eqnarray}
\sigma_{p'} &=&
\frac{ Q_1}{ \beta_1 \sqrt{s}}=\sqrt{\frac{\alpha_1 - 1}{\beta_1}} e^{i \theta_1} 
=e^{Y_{p'} + i \phi_{p'}} \, , \nonumber\\
\sigma_{q'} &=&   
\frac{ Q_2}{\left(1+ \beta_2\right) \sqrt{s}}=\sqrt{\frac{\alpha_2}{1+\beta_2}} e^{i \theta_2}=e^{Y_{q'}+ i \phi_{q'}} \, ,\\
\sigma_{k} &=& \frac{ Q_1- Q_2 }{\left(\beta_1- \beta_2 \right) \sqrt{s}} =
\frac{ \sqrt{(\alpha_1-1)\beta_1} e^{i \theta_1}- \sqrt{(1+\beta_2) \alpha_2} e^{i \theta_2} }{\beta_1- \beta_2 } = e^{Y_k + i \phi_k} \, . \nonumber
\end{eqnarray}
As it is shown in Fig.~\ref{Stereo4}, on the Riemann sphere they lie onto the equatorial plane on circumferences with radii
\begin{eqnarray}
R_{p'} = 2{\sqrt{(\alpha_{1}-1)\beta_{1} \over (1-\alpha_{1}-\beta_{1})^{2}}} 
\, \, \, , \, \, \,
R_{q'} = 2{\sqrt{\alpha_{2}(1+\beta_{2}) \over (1+\alpha_{2}+\beta_{2})^{2}}} \, ,  \nonumber\\
R_{k} = {2|Q_{1}-Q_{2}|\over \alpha_{1}+\beta_{1}-\alpha_{2}-\beta_{2}}=2\sqrt{(\alpha_{1}-\alpha_{2})(\beta_{1}-\beta_{2})\over (\alpha_{1}+\beta_{1}-\alpha_{2}-\beta_{2})^{2}} \, .
\end{eqnarray}

It is a simple task to find the second solution the SE using the Sudakov's variables 
since it essentially corresponds to the complex conjugate of Fairlie's solution. Both together can then be written as 
\begin{eqnarray}
\sigma_{p'}^{(+)} &=&  {\sigma_{p'}^{(-)*}} =\frac{{\widehat Q}_1 e^{-i \theta_2}}{\beta_1} 
=\sqrt{\frac{\alpha_1-1}{\beta_1}}  e^{i (\theta_1-\theta_2 + \pi)} \, , \nonumber \\
\sigma_{q'}^{(+)} &=& {\sigma_{q'}^{(-)*}}= \frac{{\widehat Q}_2 e^{-i \theta_2}}{1+\beta_2} 
= \sqrt{\frac{\alpha_2}{1+\beta_2}} \, , \label{eq:gen_sol_A5_sigmas} \\
\sigma_{k}^{(+)} &=& {\sigma_{k}^{(-)*}}= \frac{({\widehat Q}_1 - {\widehat Q}_2) e^{-i \theta_2}}{\beta_1 - \beta_2} ~=~ \frac{\sqrt{(\alpha_1-1)\beta_1} e^{i (\theta_1-\theta_2)}- \sqrt{\alpha_2 (1+\beta_2)} }{\beta_1-\beta_2} \, .\nonumber
\end{eqnarray} 
The 5--point amplitude for the $\varphi^{3}$ scalar theory can now be written as the sum over the two solutions:
\begin{eqnarray}
\mathcal{A}_5^{\varphi^3}
&=&\int d z_{p'} d z_{q'} \, {\cal J}^{-1}\delta\big(z_{p'}-\sigma_{p'}\big) \delta\big(z_{q'}-\sigma_{q'}\big) 
\frac{ z_{k}^2 }{z_{q'}^2 z_{q'k}^2 z_{kp'}^2 } + {\rm c.c.} \nonumber \\
&=&{2\over s^2} \mbox{Re\,} \left[\left({ \sigma_{p'} \over \sigma_{q'}} \right){ 1 \over L \widetilde{L} 
-R\widetilde{R}}\right]
\label{eq:A5_final1} \\
&=&  {1\over s^{2}}\left[\frac{1}{\alpha_1+\beta_1}  - \frac{1}{\alpha_2+\beta_2}
+\frac{1}{(\alpha_1+\beta_1) \beta_1} - \frac{1}{\beta_1 \alpha_2}
+\frac{1}{\alpha_2 (\alpha_2+\beta_2) }\right] \, ,
\nonumber 
\end{eqnarray}
where
\begin{eqnarray}
L &=& {\sigma_{p'k}\over \sigma_{p'q'}} \left[(\alpha_1 - 1) \frac{\sigma_{q'}}{\sigma_{p'}} + \beta_1 \frac{\sigma_{p'}}{\sigma_{q'}}\right] \, ,\nonumber \\
R &=& \left({\sigma_{p'}\sigma_{p'k}\over \sigma_{p'q'}}\right)
{(1-\alpha_1+\alpha_2-\beta_1+\beta_2) (\alpha_1+\beta_1)\over 
(\alpha_1 - \alpha_2 + \beta_1)\sigma_{p'} - (1 + \beta_2) \sigma_{q'}} \, ,
\end{eqnarray}
and the quantities with tilde are defined by 
\begin{eqnarray}
\widetilde{\cal O}\Big(\alpha_1,\alpha_2,\beta_1,\beta_2,\theta_{1}-\theta_{2}\Big) =  {\cal O}\Big(1-\alpha_2, 1-\alpha_1,-1-\beta_2,-1-\beta_1,\theta_{2}-\theta_{1}\Big).
\end{eqnarray}
This is a first  analysis of the use of Sudakov variables in the context of the CHY calculation of scattering amplitudes. In the future it will be interesting to generalize this representation to $n$-point amplitudes.
\\
\\
The research here presented has been supported by the Spanish Research Agency (Agencia Estatal de Investigaci\'on) through the grant IFT Centro de Excelencia Severo Ochoa SEV-2016-0597, and the Spanish Government grants FPA2015-65480-P, FPA2016-78022-P.

\printindex

\end{document}